\documentclass[]{aa}  
\usepackage{amsmath}
\usepackage[colorlinks,breaklinks]{hyperref}
\hypersetup{linkcolor=blue,citecolor=blue,filecolor=black,urlcolor=blue}
\usepackage{xcolor}
\usepackage{amssymb}

\newcommand\uncIa{`Ia-unclear'}
\newcommand{\kms}{km~s$^{-1}$}

\newcommand{\SiII} {\ion{Si}{ii}}
\newcommand{\SiIIs} {\SiII\ $\lambda6355$}
\newcommand{\SiIIf} {\SiII\ $\lambda5972$}
\newcommand{\Msun}{\ensuremath{{M}_\odot}}
\newcommand{\Mstar}{\ensuremath{{M}_\star}}
\newcommand{\angstrom}{\mbox{\normalfont\AA} }
\newcommand{\md}[1]{{\color{red} #1}}

\usepackage{xspace}
\usepackage{graphicx}
\usepackage{txfonts}
\usepackage{pdflscape}
\usepackage{comment}
\usepackage{orcidlink}

\newcommand{\customcitecolor}[2]{%
    \begingroup
    \hypersetup{citecolor=#1}%
    \cite{#2}%
    \endgroup
}

\newcommand{\customcitepcolor}[2]{%
    \begingroup
    \hypersetup{citecolor=#1}%
    \citep{#2}%
    \endgroup
}

\begin{document}

%
%
   \title{ZTF SN~Ia DR2: Properties of the low-mass host galaxies of Type Ia supernovae in a volume-limited sample}
      \author{U. Burgaz \inst{1}\orcidlink{0000-0003-0126-3999}
    \and K. Maguire\inst{1}\orcidlink{0000-0002-9770-3508}
    \and G. Dimitriadis\inst{1,2}\orcidlink{0000-0001-9494-179X}
    \and M.~Smith\inst{2}\orcidlink{0000-0002-3321-1432}
    \and J. Sollerman\inst{3}\orcidlink{0000-0003-1546-6615}
    \and L.~Galbany\inst{4,5}\orcidlink{0000-0002-1296-6887}
    \and M.~Rigault\inst{6}\orcidlink{0000-0002-8121-2560}
    \and A.~Goobar\inst{7}\orcidlink{0000-0002-4163-4996}
    \and J.~Johansson\inst{7}\orcidlink{0000-0001-5975-290X}
    \and Y.-L.~Kim\inst{2}\orcidlink{0000-0002-1031-0796}  
    \and A.~Alburai\inst{4,5}\orcidlink{0009-0007-2731-5562}
    \and M.~Amenouche\inst{8}\orcidlink{0009-0006-7454-3579} 
    \and M.~Deckers\inst{1}\orcidlink{0000-0001-8857-9843}
    \and M.~Ginolin\inst{6}\orcidlink{0009-0004-5311-9301}
    \and L. Harvey\inst{1}\orcidlink{0000-0003-3393-9383}
    \and T.~E.~Muller-Bravo\inst{4,5}\orcidlink{0000-0003-3939-7167}
    \and J.~Nordin\inst{9}\orcidlink{0000-0001-8342-6274}
    \and K.~Phan\inst{4,5}\orcidlink{0000-0001-6383-860X}
    \and P.~Rosnet\inst{10}\orcidlink{0000-0002-6099-7565}  
    \and P.~E.~Nugent\inst{11,12}\orcidlink{000-0002-3389-0586}
    \and J.~H.~Terwel\inst{1,13}\orcidlink{0000-0001-9834-3439}
    \and M.~Graham\inst{14} \orcidlink{0000-0002-3168-0139}
    \and D.~Hale\inst{15}
    \and M.~M.~Kasliwal\inst{14}\orcidlink{0000-0002-5619-4938}
    \and R.~R.~Laher\inst{16}\orcidlink{0000-0003-2451-5482}
    \and J.~D.~Neill\inst{14}\orcidlink{0000-0002-0466-1119}
    \and J.~Purdum\inst{15}\orcidlink{0000-0003-1227-3738}
    \and B.~Rusholme\inst{16}\orcidlink{0000-0001-7648-4142}  
}

    \institute{School of Physics, Trinity College Dublin, College Green, Dublin 2, Ireland\\
            \email{burgazu@tcd.ie}
            \and Department of Physics, Lancaster University, Lancs LA1 4YB, UK\            
            \and Oskar Klein Centre, Department of Astronomy, Stockholm University, SE-10691 Stockholm, Sweden\
            \and Institute of Space Sciences (ICE, CSIC), Campus UAB, Carrer de Can Magrans, s/n, E-08193, Barcelona, Spain\   
            \and Institut d'Estudis Espacials de Catalunya (IEEC), E-08034 Barcelona, Spain\  
            \and Univ Lyon, Univ Claude Bernard Lyon 1, CNRS, IP2I Lyon/IN2P3, UMR 5822, F-69622, Villeurbanne, France\ 
            \and Oskar Klein Centre, Department of Physics, Stockholm University, SE-10691 Stockholm, Sweden\
            \and National Research Council of Canada, Herzberg Astronomy \& Astrophysics Research Centre, 5071 West Saanich Road, Victoria, BC V9E 2E7, Canada\
            \and Institut für Physik, Humboldt-Universit\"at zu Berlin, Newtonstr. 15, 12489 Berlin, Germany\
            \and Universit\'e Clermont Auvergne, CNRS/IN2P3, LPCA, F-63000 Clermont-Ferrand, France\
            \and Lawrence Berkeley National Laboratory, 1 Cyclotron Road, MS 50B-4206, Berkeley, CA 94720, USA\
            \and Department of Astronomy, University of California, Berkeley, 501 Campbell Hall, Berkeley, CA 94720, USA\  
            \and Nordic Optical Telescope, Rambla Jos\'e Ana Fern\'andez P\'erez 7, ES-38711 Bre\~na Baja, Spain\
            \and Division of Physics, Mathematics and Astronomy, California Institute of Technology, Pasadena, CA 91125, USA\   
            \and Caltech Optical Observatories, California Institute of Technology, Pasadena, CA  91125\
            \and IPAC, California Institute of Technology, 1200 E. California Blvd, Pasadena, CA 91125, USA\
            }

   \date{Received 11 Oct 2024; accepted 10 Dec 2024}

\titlerunning{ZTF SN~Ia DR2: Host galaxy properties of SNe~Ia}
\authorrunning{U. Burgaz et al.}
 
  \abstract
  {In this study, we explore the characteristics of `low-mass' (log(\Mstar/\Msun) $\leq$ 8) and `intermediate-mass' (8 $<$ log(\Mstar/\Msun) $\leq$ 10) host galaxies of Type Ia supernovae (SNe~Ia) from the second data release (DR2) of the Zwicky Transient Facility survey and investigate their correlations with different sub-types of SNe~Ia. We use the photospheric velocities measured from the \SiIIs\ feature, SALT2 light-curve stretch ($x_1$) and host-galaxy properties of SNe~Ia to re-investigate the existing relationship between host galaxy mass and \SiIIs\ velocities. We also investigate sub-type preferences for host populations and show that while the more energetic and brighter 91T-like SNe~Ia tends to populate the younger host populations, 91bg-like SNe~Ia populate in the older populations. Our findings suggest High Velocity SNe~Ia (HV SNe~Ia) not only comes from the older populations but they also come from young populations as well. Therefore, while our findings can partially provide support for HV SNe~Ia relating to single degenerate progenitor models, they indicate that HV SNe~Ia other than being a different population, might be a continued distribution with different explosion mechanisms. We lastly investigate the specific rate of SNe~Ia in the volume-limited SN Ia sample of DR2 and compare with other surveys.}


   \maketitle

\section{Introduction}
\label{sec:intro}

Type Ia supernovae (SNe~Ia) are believed to be thermonuclear explosions of Carbon-Oxygen (C/O) white dwarfs \citep[WDs, ][]{hoyle1960}. They have an important role in astronomy as standardizable candles and are used to measure the accelerating expansion of the universe \citep{perlmutter1997, perlmutter1999, ries1998, ries2016, des2024}. Despite the standardizable nature of SNe~Ia, the mechanisms behind their explosions and their progenitors are still not fully understood and there are potentially multiple explosion channels and mechanisms \citep{maoz2014, maeda2016, Liu2023}. There are currently two commonly considered types of SN Ia progenitor models. In the single-degenerate (SD) model \citep{whelan1973, nomoto1997} a C/O WD accretes material from the companion star, which is either a non-degenerate main-sequence star or an evolved star, such as a red giant. The double-degenerate (DD) model involves the merging of two C/O WDs \citep{iben1984, webbink1984}. There are several evolutionary paths within the DD model that can lead to SN Ia explosions. An initial helium detonation could be triggered by the accumulation of a helium shell on top of the primary carbon-oxygen (CO) WD if the secondary WD is a He WD, which would lead to sub-Chandrasekhar-mass (\textit{M$_{Ch}$} $\lesssim$ 1.2 \Msun) double-detonation scenario \citep{fink2007, sim2010, shen2018}. Over the past decades, numerical simulations have also explored the merger of two WDs, where a carbon detonation can be directly triggered by the interaction of the debris from the secondary WD with the primary WD, lead to the violent merger scenario \citep{raskin2009, pakmor2013, sato2015, roy2022}.

Several studies have tried to identify SN~Ia progenitors and improve their cosmological use by examining the connections between the photometric (e.g., absolute magnitude, light curve width, colour, Hubble residuals) and their host galaxy properties, such as morphological type \citep{hamuy1996}, offset from the host galaxy centre \citep{galbany2012},   color \citep{Kelsey2021}, star formation rate (SFR) and age \citep[][]{gallagher2005, howell2009,rigault2013,pan2014,barkhudaryan2019,hakobyan2020,Rigault2020,wiseman2023}, stellar mass \citep{sullivan2010, gupta2011, childress2013,pan2014}, and metallicity \citep{gallagher2008,pan2014}.  After applying standard corrections for light curve width and color, it is seen that SNe~Ia appear brighter in galaxies with higher stellar masses compared to those in galaxies with lower stellar masses. SNe~Ia with faster declining light curves prefer higher mass galaxies. It is also shown that galaxies with higher rates of star formation generally host brighter, slower SNe~Ia than those with lower star formation activity as seen in passive galaxies.

The spectral properties of SNe Ia have also been suggested to correlate with their host galaxy properties.
\citet{wang2013} found that SNe~Ia with \SiIIs\ absorption feature velocities at maximum light $\gtrsim$ 12,000 km s$^{-1}$ are more likely to be found more centrally in their host galaxies, while SNe~Ia with \SiIIs\ velocities of $<$ ~12,000 km s$^{-1}$ occur across a broader range of radial distances. Results from \citet{pan2015} indicates a trend amongst the HV and NV SNe~Ia, where more massive galaxies host more SNe~Ia with higher \SiIIs\ velocities than SNe Ia with lower \SiIIs\ velocities, which is further supported by future analysis, where they suggest that the main criterion for forming a HV SNe~Ia is the metallicity of the host \citep[][hereafter P20]{pan2020}.

A trend between SN Ia rates and host galaxy stellar mass and star formation rate is observed in several studies \citep[e.g.,][]{sullivan2006, li2021_rate, smith2012, graur2013, wiseman2021}, where it has been shown that the specific SN~Ia rate, which is defined as the total SN~Ia rate scaled by the stellar mass of the galaxy, is considerably higher for lower mass more actively star-forming galaxies than for higher mass less actively star-forming galaxies. The delay-time distribution (DTD) is the rate of explosion of a SN Ia as a function of the time since a burst of star formation, and can provide constraints on the explosion mechanism. For example, SD scenarios struggle to match the observed rate of SNe Ia at long delay times, while DD scenarios under-predict the rate at short delay times \citep{wangbo2012,maoz2014, Liu2023}. 

\citet{brown2019} studied the specific rate of low-redshift SNe Ia from the 
All-Sky Automated Survey for Supernovae \citep[ASAS-SN;][]{shappee2014,kochanek2017} sample and by using the stellar mass function (SMF) presented in \cite{Bell2003}, showed that the specific SN Ia rate is roughly proportional to the inverse square root of stellar mass. \cite{gandhi2022} presented a modified version of the ASASS-SN sample by converting the \cite{Bell2003} SMF to the SMF of \cite{baldry2012}, where they also used the measurements in the \citep{wiseman2021} coming from the the Dark Energy Survey \citep[DES;][]{des2016} and investigated the SMF dependence of the specific SN Ia rates. 

\cite{kistler2013} argued that the specific SN Ia rate's dependence on stellar mass could be influenced by metallicity as lower-mass galaxies typically contain lower-metallicity stars where these lower-metallicity stars produce higher-mass white dwarfs, which are more prone to explosion than their higher metallicity counterparts. \cite{gandhi2022} utilized FIRE$-$2 simulations to show that incorporating a SN Ia rate that scales inversely with metallicity ($Z^{-0.5}$ to $Z^{-1}$) enhances the match between simulated galactic stellar masses and observations.

It is known that a minor fraction of SNe takes place in either dwarf galaxies or globular clusters that are too dim to be observed, hence considered as `hostless'. While this case more commonly seen in super-luminous SNe \citep{galyam2012} since they tend to appear in faint dwarf galaxies, a small number of spectroscopically normal SNe Ia also appear hostless \citep{graham2015,moon2021} or happen in very faint dwarf galaxies \citep{holoien2023}. Restrictions on identifying and matching host galaxies could lead to falsely classifying SNe as hostless in large datasets (see Section \ref{sec:sec2galaxyphot} for further details on host galaxy identification and matching for the ZTF sample). In this study, we account for all these factors, identify the hostless SNe in our sample that are actually located in faint-undetected galaxies, and include them in our analysis.

In this paper, we present an analysis of the host galaxies from the Zwicky Transient Facility (ZTF) data release (DR2) SN Ia, obtained as part of the Zwicky Transient Facility survey from 2018 - 2020 \citep[][]{bellm2019, graham2019, masci2019, dekany2020}. ZTF is an optical time-domain survey that has been in operation since 2018.  The ZTF SN Ia DR2 sample contains 3628 spectroscopically confirmed SNe Ia (see \customcitecolor{red}{Rigault2024a} and \md{Smith et al. (in preparation)}), without biases in terms of host-galaxy properties (i.e.~massive galaxies were not specifically targeted). Most of the earlier studies are largely biased towards higher mass (higher luminosity) galaxies since they are easier to observe. Hence, missing a considerable fraction of low-mass host galaxies. We focus on SNe Ia in low mass hosts, including those without detected hosts (`hostless' SNe Ia). We discuss the broad sample selection, hostless SNe~Ia, and the low-mass galaxies in our sample in Section~\ref{sec:data_analysis}. We present the results from our study and compare our results with previous host galaxy studies in Section~\ref{sec:results}. Conclusions are presented in Section~\ref{sec:conclusions}.

\section{Data and analysis}
\label{sec:data_analysis}

In this section, we discuss the data selection and analysis of our low-mass host galaxy sample from the ZTF SN Ia DR2. Section~\ref{sec:SNiadata} describes our low-mass sample selection based on photometric and spectroscopic properties, while Section~\ref{sec:sec2galaxyphot} details the host galaxy selection criteria. In Section~\ref{sec:hostless}, we present the hostless SNe Ia from the ZTF SN Ia DR2 sample.

\begin{figure*}
\centering
\includegraphics[width=\linewidth]{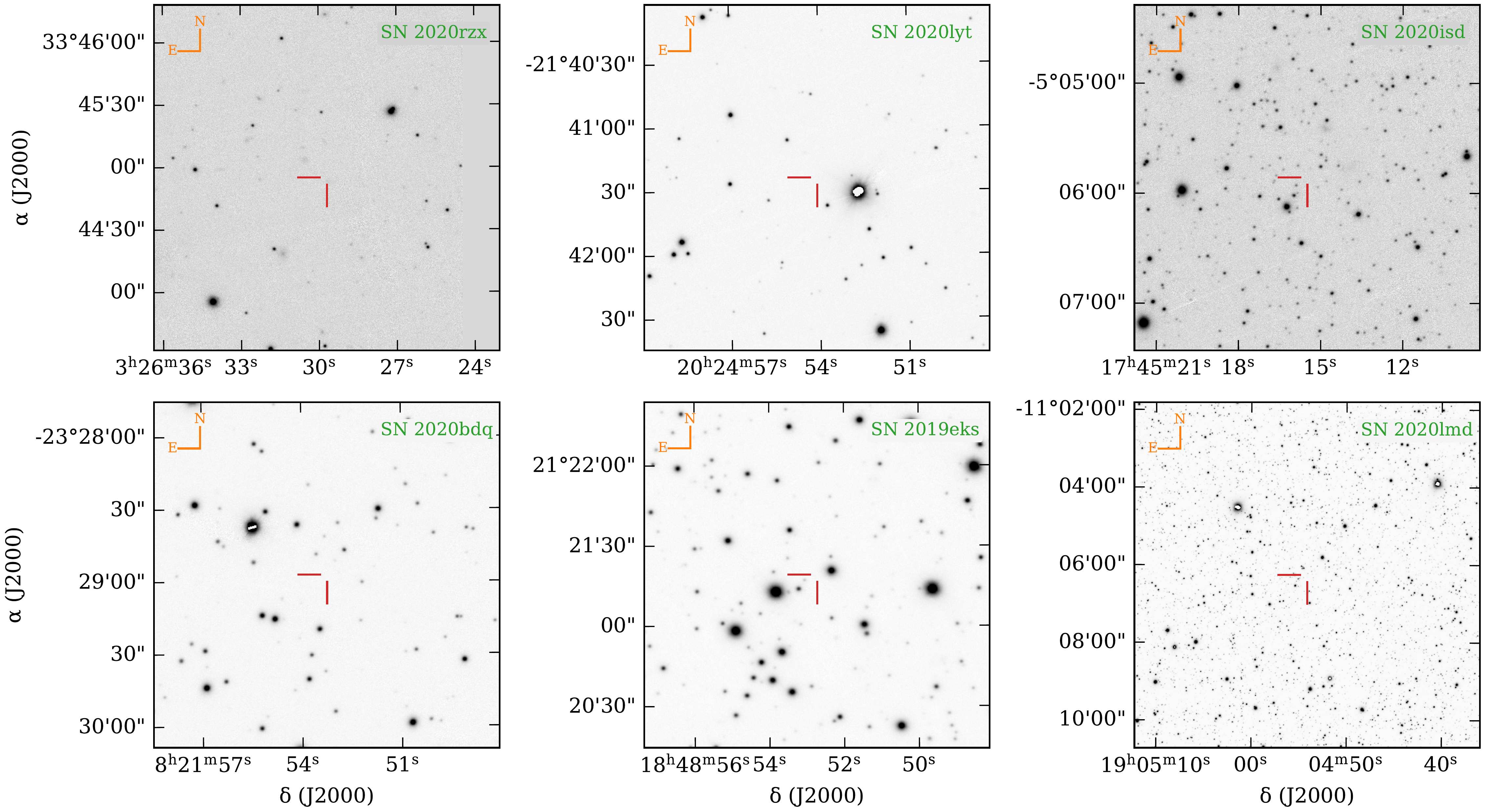}
\caption{Hostless SNe~Ia from the volume-limited ZTF DR2 sample. 
All SNe presented in this plot are investigated for potential host galaxies within a search radius of 125 kpc and each image shown here spans 125 kpc scaled from the corresponding SN redshift for a clear visualization.} 
\label{fig:hostless}
\end{figure*}

\subsection{ZTF DR2 SN Ia data}
\label{sec:SNiadata}

\begin{table}
\centering
\caption{Summary of the cuts applied to obtain the final sample.}
\label{tab:cuts}
\begin{tabular}{l c c}
\hline\\[-0.5em]
     & Criteria & No. of SNe \\[0.15em]
    \hline\\[-0.8em]
    \hline\\[-0.5em]
ZTF SN Ia DR2 & & 3628\\[0.30em]
Redshift cut & $z$ $\leq$ 0.06 & 1584\\[0.30em]
SN Ia type cut* & & 1535\\[0.30em]
Unreliable host cut & & 1523\\[0.30em]
LC coverage cut& & 1106\\[0.30em]
Phase cut & $-$5 d $\leq$ $t_0$ $\leq$ 5 d & 477\\[0.30em]
    \hline\\[-0.5em]  
\end{tabular}
 \begin{flushleft}
*In this study all peculiar types are excluded (see Section \ref{sec:SNiadata}).
 \end{flushleft}
\end{table}

3628 spectroscopically confirmed SNe~Ia events are released as part of the ZTF DR2 sample (see \customcitecolor{red}{Rigault2024a} and \md{Smith et al. (in preparation)}). We make a number of cuts to this sample to obtain our final sample for analysis, which are detailed in Table \ref{tab:cuts} to show how we arrive at our final sample. The ZTF DR2 SNe Ia sample is volume-limited and unbiased up to redshift ($z$) $\leq$ 0.06, with key parameters like stretch, color, and peak magnitude accurately reflecting their intrinsic distributions. Simulations by \customcitecolor{red}{Amenouche2024} demonstrate that the sample is complete for non-peculiar SNe Ia at redshift ($z$) $\leq$ 0.06, which is supported by the color distribution \customcitepcolor{red}{Ginolin2024b} that aligns with the intrinsic properties of the population, as confirmed by KS tests with approximately 2-sigma confidence. Hence we only consider this spectroscopically-complete sample of SNe~Ia in this work. This cut is made to ensure that we can accurately determine the rates of SNe Ia across different galaxy types.

The next cut we apply is remove SNe Ia that are spectroscopically classified as peculiar. The spectral classifications used to subtype the SNe Ia in this analysis come from two sources. For the SNe Ia with peak ($-$5 d $\leq$ $t_0$ $\leq$ 5 d) spectra, where $t_0$ is the time of peak, we used the subtypes of \customcitecolor{red}{burgaz2024} and for the rest of the sample we take the subtypes from \customcitecolor{red}{Dimitriadis2024}. While both subtype classifications are self consistent, \customcitecolor{red}{Dimitriadis2024} lacks 04gs-like and 86G-like classifications. Hence, for the spectra outside the peak range we estimate we could have approx.~13 04gs-like and 14 86G-like SNe~Ia (estimated from scaling the peak sample to the full sample) that could be misclassified as a normal SNe Ia. These classifications are based on visual inspection as well as investigation of the properties of their main spectral features, including velocity, pseudo-equivalent width (pEW).  The peculiar subtypes that are excluded include `Ia-CSM', `Iax', `03fg-like', `02es-like', `00cx-like' and `18byg-like' SNe~Ia. This removes 49 SNe Ia from our sample. We still consider 91T-like, 99aa-like, 04gs-like, 86G-like and 91bg-like subtypes \customcitepcolor{red}{burgaz2024}. `Ia-unclear' subtype refers to SNe~Ia where detailed investigations did not reveal a confident subtype, and we see a similar distribution in each mass bin in the low mass regions. This subtype also includes those classified after post peak brightness.

There are 18 SNe Ia in the ZTF DR2 sample for which the host galaxy masses have not been calculated. Eight have potential host galaxies with higher Directional Light Radius \citep[$d_{DLR}$;][]{sullivan2006,smith2012,gupta2016} matches, ranging from 7.1 to 14, and were thus excluded since the criterion to match a host galaxy was to have a $d_{DLR}$ of less than 7 \md{(Smith et al.~in prep.)}. Five are located in clusters, making it impossible to determine the actual host galaxy. Detailed host galaxy associations are discussed briefly in Section~\ref{sec:sec2galaxyphot} and in detail in \md{Smith et al.~(in prep.)}. While we discuss about the six truly hostless SNe~Ia in Section~\ref{sec:hostless}, in this work, we exclude the remaining 12 SNe~Ia (corresponding to less than 1 per cent of the sample) from our studies. This results in a final sample of 1523 SNe Ia that we consider in this work.

In some cases, we wish to compare the host galaxy properties to their light curve parameters. Therefore, we also can apply a cut based on the quality of the light curves so that we have a sub-sample with reliable \texttt{SALT2} light curve \citep{guy2007} parameters. These cuts follow those of the DR2 sample \customcitepcolor{red}{Rigault2024a}, which require each SN to have a minimum of two detections in two different filters (both pre- and post-maximum light relative to the time of maximum light,  \textit{t$_0$}) and a total of at least seven detections across all filters. In this work, we use the light curve parameters ($x_1$, $c$) from \md{Smith et al.~(in prep.)}, in which the values were calculated with the \texttt{SALT2} version published by \citet{taylor2021}, only considering the -10 to +40 days in the rest-frame phase range as suggested by\customcitepcolor{red}{Rigault2024b}. While some of the analysis in this work does not require a phase cut, when we compare the host galaxy properties to their spectroscopic properties we use a phase cut of $-$5 d $\leq$ $t_0$ $\leq$ 5 d to study the sample at peak brightness. Table~\ref{tab:cuts} shows all the criteria used to define the final sample along with the number of remaining SNe~Ia after each cut. 

The velocities and pEWs of the \SiIIf\ and \SiIIs\ features for the volume-limited ZTF DR2 SN Ia sample in maximum light ($-$5 d $\leq$ $t_0$ $\leq$ 5 d) are calculated and presented in \customcitecolor{red}{burgaz2024}. The majority of the optical spectra comes from the low-resolution integral field spectrograph, SED machine \cite[SEDm,][]{blagorodnova2018, rigault2019, kim2022} on the 60-inch telescope at the Mount Palomar observatory. A detailed methodology and selection criteria for the sample is presented in \customcitecolor{red}{burgaz2024}. 

In the ZTF DR2 SN Ia sample, the redshifts come from three different sources, i) a catalogue redshift, where the redshift of the galaxy is known from SIMBAD, NASA extra-galactic database (NED), Sloan Digital Sky Survey (SDSS) DR 17 \citep{Abdurrouf2022} or the Dark Energy Spectroscopic Instrument (DESI) through the Mosthost program \citep{Soumagnac2024}, ii) galaxy features in the SN spectra, or iii) where no redshift information exists, template matching \textsc{snid} is used to estimate the redshift. 

\subsection{ZTF DR2 host galaxy data}
\label{sec:sec2galaxyphot}

The method used to identify the host galaxies of the ZTF SN Ia DR2 sample, along with how the host galaxy properties are calculated is presented in \md{Smith et al.~(in prep.)}. Briefly, the first step was to obtain the 100 kpc radius images (centered on the SN location) from the Legacy Imaging Survey DR9 \citep[LS;][]{Dey2019},  SDSS DR17  and Pan-STARRS DR2 \citep[PS1;][]{Chambers2016} (in given priority order). The next step involved applying the software package \texttt{Host\_Phot} \citep{mullerbravo2022} to the PS1 images to obtain the Kron fluxes. In this process $d_{DLR}$ for each SN is also calculated. A cut off of $d_{DLR}<7$ is chosen to associate possible host galaxies, where more than one possible galaxy exist the source with smaller $d_{DLR}$ is chosen as the host galaxy. \md{Smith et al.~(in prep.)} also calculated the stellar galaxy masses from both the method of \citet{Taylor2011} and \texttt{P\'{E}GASE.2} \citep{fioc1997}. Both methods, especially in the higher masses produce similar results. The masses calculated with \texttt{P\'{E}GASE.2} is used throughout this paper.

\begin{table*}
\tiny
\caption{Properties of the hostless SNe~Ia from the volume-limited ZTF DR2 SN Ia sample.}
\begin{tabular}{l c c c c c c c c c c c}
\hline\\[-0.5em]
SN Name & Redshift$^*$ & Subtype & SALT2 $x_1$  & SALT2 $c$ & $M_g$$_{, SN}$$^3$ & $M_i$$_{, host}$ upper & log(\Mstar/\Msun) & A$_V$ & Galactic latitude ($b$)\\[0.15em]
 & & & & & (mag) & limit (mag) & upper limit & (mag) & (deg)\\[0.15em]
    \hline\\[-0.8em]
    \hline\\[-0.5em]

SN 2019eks & 0.047$\pm$0.003 & 99aa-like & 0.66$\pm$0.16 & $-$0.12$\pm$0.04 & $-$19.00$\pm$0.04 & $-$14.0 & <6.6 & 0.67 & 10.1\\[0.30em]
SN 2020bdq & 0.043$\pm$0.004 & normal & 0.07$\pm$0.40 & $-$0.15$\pm$0.04 & $-$18.90$\pm$0.06 & $-$13.6 & <6.6 & 0.37 & 7.6 \\[0.30em]
SN 2020isd & 0.030$\pm$0.003 & Ia-unclear & 1.07$\pm$0.35 & 0.13$\pm$0.04 & $-$18.98$\pm$0.06 & $-$14.2 & <6.0 & 2.18 & 12.2\\[0.30em]
SN 2020lmd & 0.011 & normal & 2.12$\pm$0.29 & 0.01$\pm$0.04 & $-$18.31$\pm$0.05 & $-$11.2 & <5.3 & 1.11 & $-$8.0\\[0.30em]
SN 2020lmd$^\dagger$ & 0.018 & normal & 2.12$\pm$0.29 & 0.01$\pm$0.04 & $-$18.31$\pm$0.05 & $-$12.1 & <6.0 & 1.11 & $-$8.0 \\[0.30em]
SN 2020lyt & 0.037$\pm$0.003 & Ia-unclear & $-$0.42$\pm$1.08 & 0.13$\pm$0.15 & $-$19.04$\pm$0.29 & $-$13.2 & <6.5 & 0.15 & $-$29.8\\[0.30em]
SN 2020rzx & 0.041$\pm$0.003 & normal & 1.29$\pm$0.09 & $-$0.07$\pm$0.03 & $-$19.18$\pm$0.04 & $-$13.7 & <6.5 & 0.64 & $-$18.8\\[0.30em]
    \hline\\[-0.5em]  
\end{tabular}\\
{\raggedright($*$)The upper bounds of the redshifts are used to place the most conservative limits on the host parameters.\par} 
{\raggedright($^2$) Absolute magnitude of the SN in ZTF $g$-band.\par} 
{\raggedright($^\dagger$) Due to the HV feature observed in the \SiIIs\ line, the SN matches better with observations at higher redshifts. In this study, we identified the upper range of redshifts where the SN showed the best match for the HV feature. In the subsequent analysis, we use the values calculated from this optimal redshift value.\par} 
\label{tab:prophostless}
\end{table*}

\subsection{Hostless SNe Ia}
\label{sec:hostless}

There are six SNe Ia in the our sample that were not associated with any host galaxy in the DR2 and we label these as `hostless'. This is an observed naming convention since each SNe Ia will occur in a host galaxy of some mass. The observational occurence of these `hostless' SNe Ia in our sample is 0.39 per cent.  We visually inspected the PS1 r-band images that spans over 125 kpc for each potentially hostless SNe Ia. Fig.~\ref{fig:hostless} shows the location and the surroundings of these SNe~Ia. No candidate galaxy could be identified for five of the six SNe Ia.  For SN 2020rzx, a diffuse source at the detection limit of PS1 can be seen at 3 kpc away from the SN location. However, there are no photometric measurements available in the PS1 database for this object. \texttt{Host\_Phot} did not return any flux from the images either.

We used the limiting magnitudes of the PS1 survey g- and i-band images to calculate absolute magnitude limits and upper limits of the stellar galaxy masses for each of the six SNe Ia. To obtain the most conservative absolute magnitudes and mass limits, we use the upper bound of the redshift uncertainty for each event as the redshift when calculating the mass. For SN 2020lmd, the redshift is quite uncertain (0.011 $\leq z \leq\ $0.018) because of the presence of a prominent high-velocity component in its \SiIIs\ feature, which results in reasonable \textsc{snid} \citep{Blondin2007} spectral template matches at a broad range of redshifts (0.011 $\leq z \leq\ $0.018), leading to an upper limit of log(\Mstar/\Msun) = 6.0. In Table~\ref{tab:prophostless} we present the photometric properties of each hostless SN Ia, along with the limits of their host galaxy properties. In terms of spectral subtyping, three of the hostless SNe are classified as normal, one as 99aa-like and remaining two as Ia-unclear, meaning that a robust sub-classification could not be made (see \customcitecolor{red}{Dimitriadis2024} and \customcitecolor{red}{burgaz2024}).

\begin{figure}
\centering
\includegraphics[width=\columnwidth]{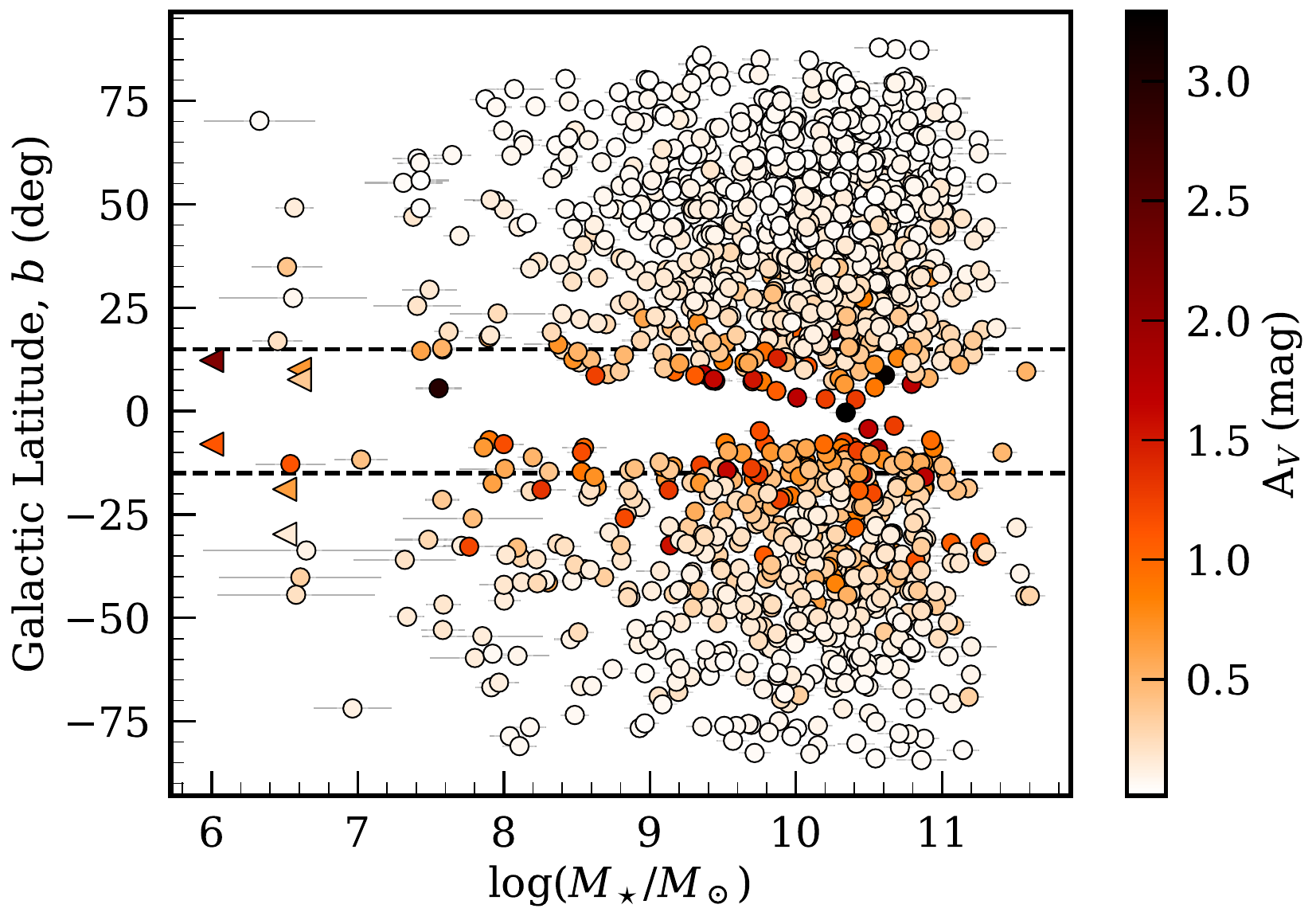}
\caption{Galactic latitudes plotted against the stellar masses from the ZTF DR2 volume limited ($z$ $\leq$ 0.06) sample, color mapped with the MW extinction (A$_V$). The filled circles represent the sample with a measured stellar galaxy mass and triangles represent the estimated upper limits of the 6 hostless SNe. Dashed black lines show +15 and -15 degrees in the Galactic latitude.} 
\label{fig:galactic}
\end{figure}

To attempt to explain why we not identify host galaxies for these relatively nearby ($z \leq $ 0.06) SNe Ia, we investigated their locations relative to the Galactic plane and their Galactic extinction values. In Fig.~\ref{fig:galactic} we show the Galactic latitude for each SN Ia in our sample against the host stellar mass, with the value of A$_V$ shown as a color bar. Four out of our six hostless SNe Ia are located within 15 degrees of the Galactic plane and exhibit high extinction values, with A$_V$ reaching up to 2.18 mag. This may push any faint host galaxy beyond the detection limit of the PS1 survey. Generally, ZTF focuses on regions of higher Galactic latitude because of the increased extinction towards the plane. 

The two remaining hostless events (SNe 2020rzx and 2020lyt) that are at 18.8 and 29.8 deg from the Galactic plane have A$_V$ values of 0.64 and 0.15 mag, respectively. SN 2020rzx is the one hostless event, where a potential source was detected in the PS1 images. Although the extinction towards SN 2020lyt is not particularly high (the mean of the ZTF is 0.23 mag), there is also a bright star located around at 21 arcsec distance from SN 2020lyt, which also could potentially conceal a faint host galaxy. Therefore, we conclude that relatively high values of extinction is the most plausible reason for no host being detected for these SNe Ia. We include these SNe Ia in our further analysis but with upper limits on their host galaxy properties.

\begin{figure}
\centering
\includegraphics[width=\columnwidth]{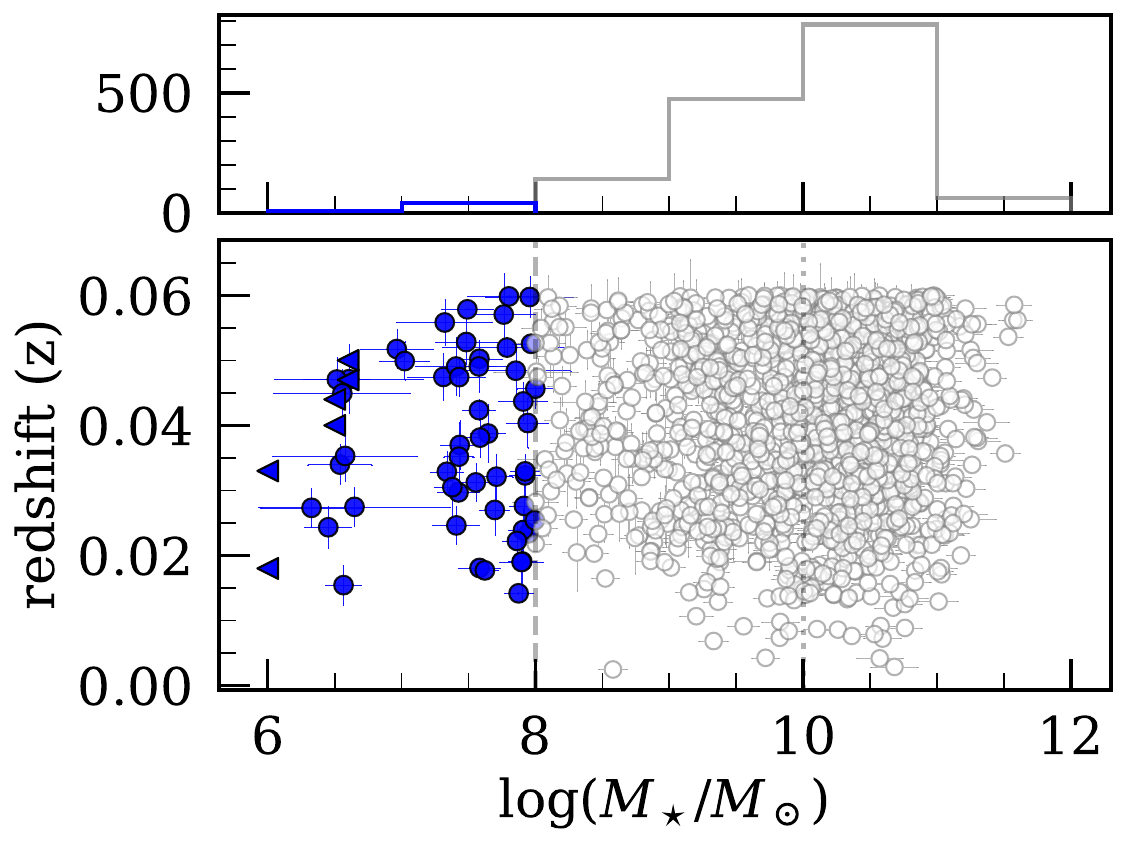}
\caption{Host galaxy masses in the ZTF DR2 volume limited ($z$ $\leq$ 0.06) sample. Blue solid circles are the `low-mass' galaxies where log(\Mstar/\Msun) $\leq$ 8. Blue triangles represent the estimated upper limits of the 6 hostless SNe. The upper histogram shows the distributions of the stellar galaxy masses in mass bins.} 
\label{fig:logmhist}
\end{figure}

\begin{figure*}
\centering
\includegraphics[width=16cm]{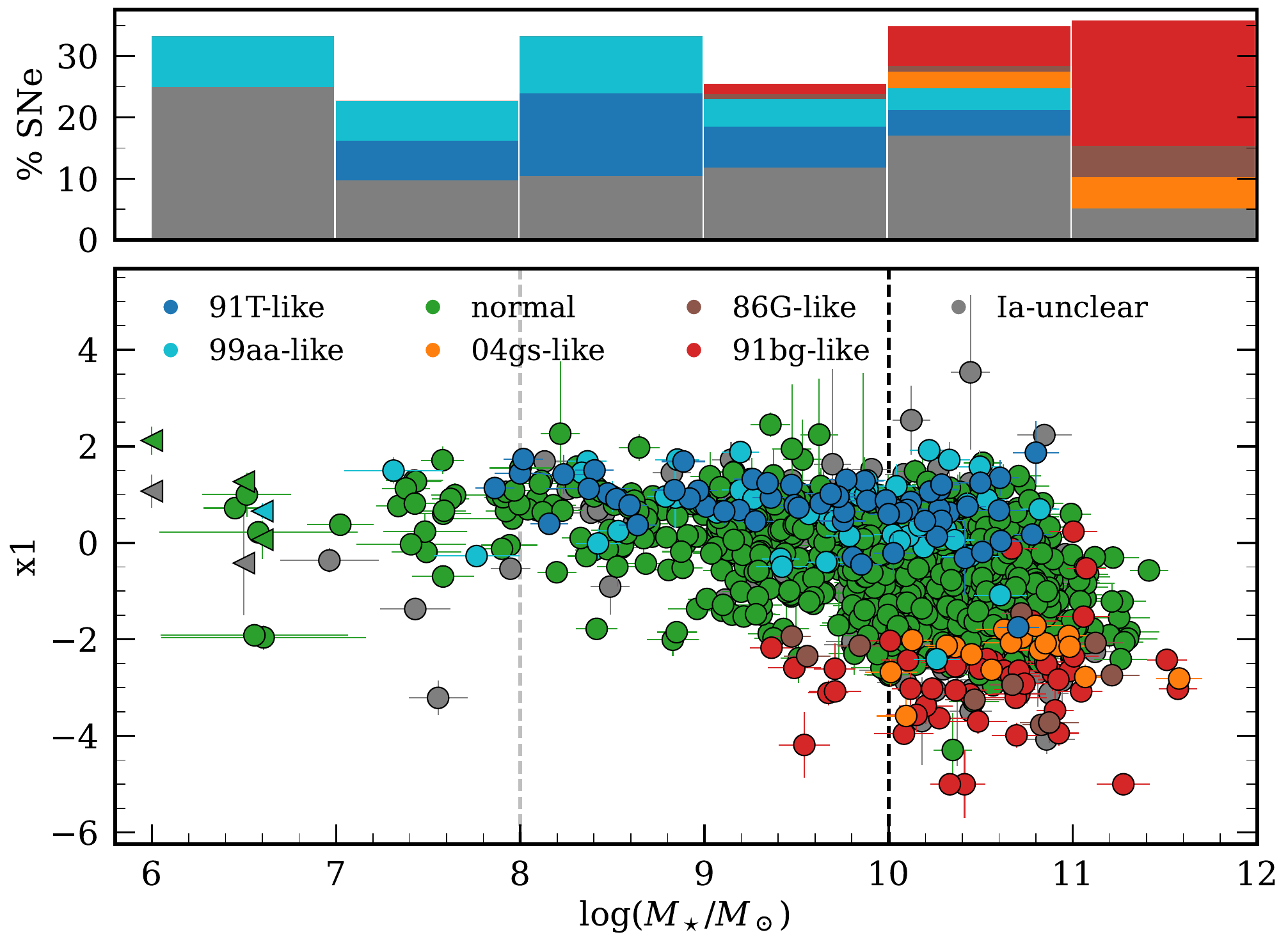}
\caption{Stellar galaxy masses plotted against the SALT2 $x_1$ parameter for the good light-curve coverage sample of the volume-limited sample of ZTF DR 2 SN Ia (1106 SNe~Ia, including the 6 hostless SNe~Ia). A relative percentage of subtypes in each mass bin is plotted as a histogram above the figure and the cumulative occurrence of these types is illustrated as additive. The distribution of the subtype normal is excluded where it is mostly constant in each mass bin with a range of (65-77 per cent). Different colors of circles, represent different subtypes used in this study. For the selection of subtype methods see Sec.~\ref{sec:x1_mass}. The gray and black dashed line represents the separation for the low, intermediate and high stellar galaxy masses.} 
\label{fig:x1mass}
\end{figure*}

\section{Results and discussions}
\label{sec:results}
In Section~\ref{sec:x1_mass}, we present the connection between the galaxy masses of SNe~Ia and their SALT2 light curve properties. We investigate the relation between the \SiIIs\ velocities of SNe~Ia and host galaxy masses in Section~\ref{sec:vsil_mass}. In Section~\ref{sec:comppan} we compare the properties of our peak sample to the sample from \citet{pan2020}, which used the parent sample of \citet{pan2015}, and present our findings. In Section~\ref{sec:rates}, we present the specific SN Ia rates and compare the specific ZTF DR2 SN Ia rate with previous studies.

\subsection{Low mass galaxies of ZTF DR2 SN Ia}
Investigation of 1523 SNe~Ia in the ZTF DR2 that pass our cuts (Table \ref{tab:cuts}) results in 52 SNe Ia with log(\Mstar/\Msun) $\leq$ 8, along with the six `hostless' SNe Ia for which upper limits exist (Table \ref{tab:prophostless}). Fig.~\ref{fig:logmhist} shows the mass distribution of the volume-limited ZTF DR2 sample, highlighting the 52 low mass host galaxies (log(\Mstar/\Msun) $\leq$ 8) and the six upper limits for hostless events. This reveals that the occurrence rate of SN Ia in low mass galaxies, including the the hostless ones, is $\sim$3 per cent, making the ZTF DR2 sample the largest collection of low mass galaxies with spectroscopically confirmed SN Ia in the literature. \citet{brown2019} investigated 476 SNe Ia host galaxies from the All-Sky Automated Survey for Supernova (ASAS-SN) Bright Supernova Catalogues where in their sample the occurrence rate of SN Ia in low mass galaxies is also $\sim$3 per cent, representing 13 galaxies.

In our sample, 37 out of 52 SNe~Ia in the low mass galaxies have good light curve coverage, hence reliable light-curve properties such as SALT2 $x_1$ and $c$ are obtained. 16 out 52 SNe~Ia in the `low-mass' galaxies also have peak ($-$5 d $\leq$ $t_0$ $\leq$ 5 d) spectral coverage.

In the low mass region (log(\Mstar/\Msun) $\leq$ 8), the population is mainly dominated by normal SN Ia (38 SN Ia), along with three 91T-like SN Ia, two 99aa-like SN Ia, and nine Ia-unclear cases, where the available spectra are mostly only late-time (more than 15 days after maximum light) and five of the unclear cases are without good light curve coverage. There are no sub-luminous subtypes (04gs-like, 86G-like or 91bg-like) found in this lowest mass bin. Below log(\Mstar/\Msun) of 7, there are 16 events (including six hostless events), of which 10 are normal SNe Ia, one 99aa-like SN Ia and five Ia-unclear.

Previous studies have found a preference for 91T-like events to occur in star-forming host galaxies \citep{hamuy1996}. There is a higher percentage of overluminous 91T-like and 99aa-like events in the low and intermediate mass bins in our sample relative to the total SN Ia sample. However, there are 91T-like and 99a-like SNe Ia found with log(\Mstar/\Msun) of 6.5 to 11. The total number of SNe Ia in the log(\Mstar/\Msun) $\leq$ 7 mass bin is low overall so the lack of any of these events is likely due to low number statistics but in the highest mass bin (log(\Mstar/\Msun) $\geq$ 11), there appears to be an intrinsic preference for these sub-classes of SNe Ia not to occur. 

\subsection{Connection with SALT2 light curve properties}
\label{sec:x1_mass}

In Fig.~\ref{fig:x1mass}, we compare the SALT2 light-curve parameter $x_1$ against the host galaxy stellar mass, color-coded by the spectral classification of each SN Ia. The relative percentage of subtypes in each mass bin is also shown as a histogram. We split the sample into three broad mass regions, `low mass' (log(\Mstar/\Msun) $\leq$ 8), `intermediate mass' (8 $<\ $log(\Mstar/\Msun) $\leq$ 10), and `high mass' (log(\Mstar/\Msun) > 10). 

As expected, the 91T-like/99aa-like SNe Ia mainly have high $x_1$ values, indicating a slower decline rate in their light-curve evolution. However, two 99aa-like SNe~Ia, SN 2020hll (ZTF18aarikzk) and SN 2020adka (ZTF20acywefl), and one 91T-like SN Ia, SN~2020ctr (ZTF19aaklqod) show lower $x_1$ values than typically seen for these sub-classes with values of $-$2.4$\pm$0.1, $-$1.1$\pm$0.1, and $-$1.7$\pm$0.1, respectively. The weighted means of $x_1$ for 91T-like SNe Ia and 99aa-like SNe Ia in our sample (excluding the three unusual cases) are 0.72$\pm$0.06 and 0.35$\pm$0.10, respectively. 

SN 2020hll and SN 2020adka are classified as a 99aa-like event based on their \SiIIs\ pEW values of $\sim$50 and 53 \angstrom around peak light, respectively. However, SN 2020hll happened very close to the host galaxy center (d$_{DLR}$ = 0.16) and the quite reddened spectra of SN 2020adka is also dominated by galaxy light. The 91T-like SN is also heavily effected by host light (d$_{DLR}$ = 0.1). The impact of host galaxy light on SN spectra, discussed in \customcitecolor{red}{burgaz2024}, highlights that it can lead to lower pEW values and potentially incorrect sub-typing, where typical increases in pEW for the 99aa-like SNe would be around 10 to 20 \angstrom, which could push the classification of these SNe to normal.

\begin{figure*}
\centering
\includegraphics[width=15cm]{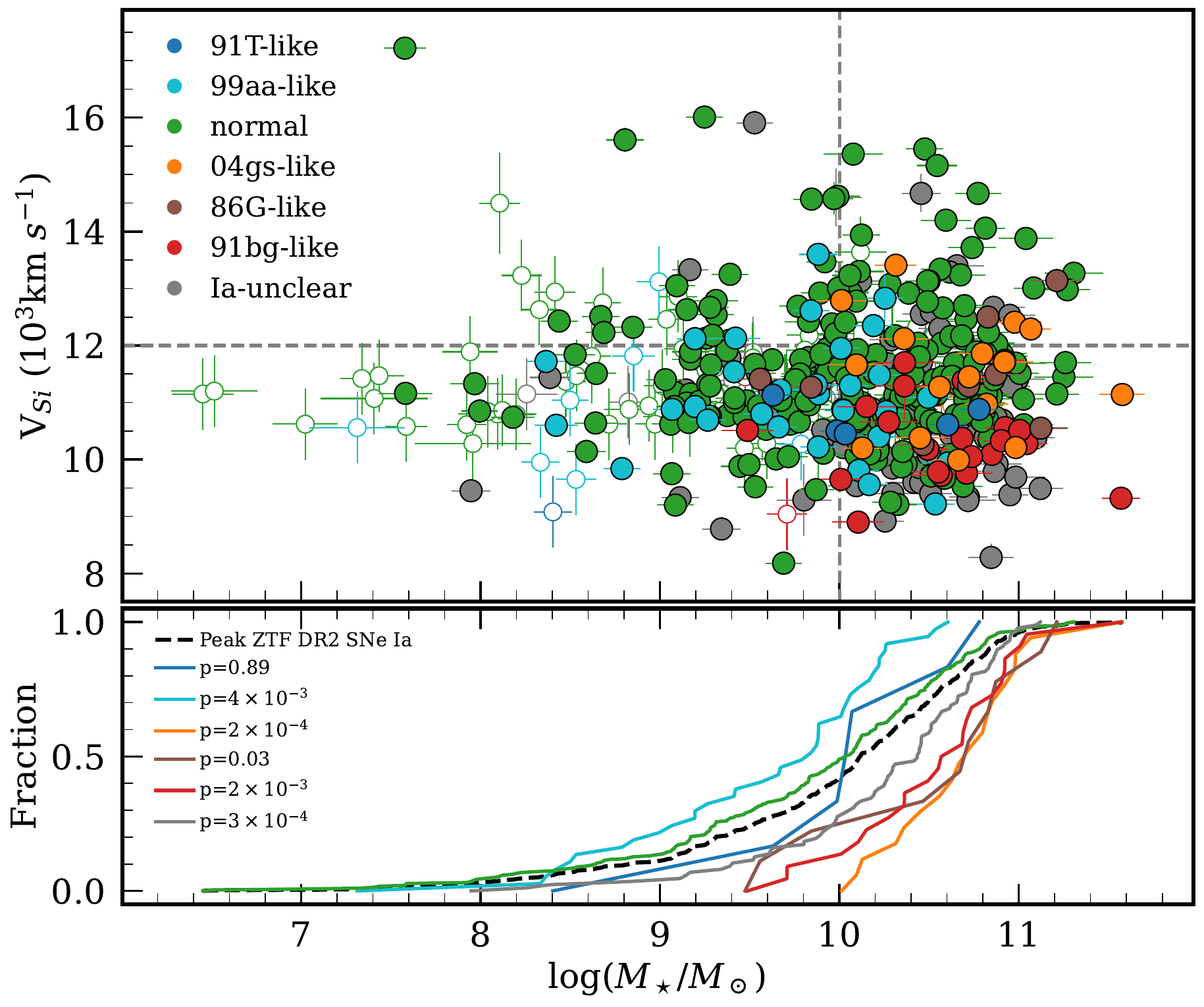}
\caption{\SiIIs\ velocities, taken from \citet{burgaz2024} plotted against the stellar galaxy masses of the volume-limited ZTF DR2 SN Ia sample for 477 SNe~Ia spectra with a phase range of $-$5 d $\leq$ $t_0$ $\leq$ 5 d. Filled and empty circles represents the SNe~Ia with known redshift sources and \textsc{snid} template matched sources, respectively. The horizontal black dashed line represents the value of \SiIIs\ velocity used to separate the sample into HV and NV events for the normal SNe~Ia (HV/NV can only be defined for normal SNe Ia). The vertical dashed line represents the standard value (10$^{10}$ \Msun) for separating high- and low-mass host galaxies. The bottom histograms show the cumulative fractions of the host galaxy masses for each subtype and the full peak sample. The p-values shown are derived from two-sample K-S tests using normal SNe Ia as the reference.} 
\label{fig:silvelmass}
\end{figure*}

The percentage of `normal' SNe~Ia in each mass bin of Fig.~\ref{fig:x1mass} is similar (in the range of 65$-$77 per cent). For the high mass bins we notice an increase in the Ia-unclear subtype and this is potentially due to the galaxies being brighter and SN Ia light being suppressed as discussed in detail in \customcitecolor{red}{burgaz2024}, where it is clearly seen both galaxy light in SN Ia spectra and the distance of the explosion to the host galaxy location directly effects the subtyping. 

On the faint end of normal SN Ia such as 04gs-like, transitional 86G-like and underluminous 91bg-like SN Ia are observed to preferentially occur in higher mass galaxies, with none observed in a galaxy below log(\Mstar/\Msun) of 9. This preference for higher mass hosts is in agreement with previous studies \citep{howell2001, sullivan2006, wiseman2021}. The mean stellar galaxy masses of 04gs-like, 86G-like and 91bg-like objects in our sample are 10.66$\pm$0.09, 10.48$\pm$0.19 and 10.51$\pm$0.10, respectively. We see a clear increase in the percentage of 04gs-like, 86G-like, and 91bg-like events with increasing stellar galaxy mass, with $\sim$30 per cent of SNe Ia occurring in galaxies with log(\Mstar/\Msun) $\geq$11 being of these sub-classes. Three 91bg-like events, SN 2020abmg (ZTF20acuoacn), SN 2018hls (ZTF18acefgoc) and SN 2018bef (ZTF18aaodnxt) in our sample show higher $x_1$ values than typically seen for this sub-class with values of $-$0.5$\pm$0.2, $-$0.1$\pm$0.2, 0.2$\pm$0.2. The weighted mean of $x_1$ for 91bg-like SNe Ia Ia in our sample (excluding the three unusual cases) is $-$2.37$\pm$0.12. The $x_1$ and stellar galaxy mass relationship for non-peculiar SNe Ia is studied in detail in \customcitecolor{red}{ginolin2024a}.

The top panel of Fig.~\ref{fig:x1mass} illustrates the relative percentages of different subtypes within each mass bin for the sample with good light-curve coverage (comprising 1106 SNe Ia). To assess whether the light-curve coverage cut influences the distribution of subtypes, we also analyzed the full sample of 1523 SNe Ia. The resulting distributions were found to be similar, agreeing within the uncertainties.

\subsection{Connection with Si II\ velocity}
\label{sec:vsil_mass}
In this section, we investigate the relation between the diversity of the velocity of the \SiIIs\ feature for SNe Ia exploding in different galaxy environments. In Fig.~\ref{fig:silvelmass}, we show the \SiIIs\ velocities against the galaxy stellar masses for the SNe Ia in our sample, separated into spectroscopic or template-matched redshift sources (see Section \ref{sec:SNiadata}). These events are limited to those around maximum light ($-$5 to 5 d) when a comparison can be made between their velocities. A significant portion (69 per cent) of SNe Ia in `low-mass' galaxies in our peak light sample do not have spectroscopic redshifts, consistent with the observational challenges posed by these fainter systems, which make them more difficult to observe and obtain reliable spectroscopic measurements. In intermediate-mass galaxies, the proportion of SNe without spectroscopic redshifts drops to 33 per cent, and in high-mass galaxies, this ratio decreases further to 9 per cent.

A classification scheme mainly based on the \SiIIs\ velocity was introduced by \cite{wang2009} by investigating the velocity distribution of `normal' SNe~Ia, where SNe with $v_{\mathrm{Si}} <$ ~12,000 km s$^{-1}$ are classified as Normal Velocity (NV) SN Ia and SNe with $v_{\mathrm{Si}} \gtrsim$ ~12,000 km s$^{-1}$ are classified as High Velocity (HV) SN Ia. We note that this classification only applies to normal SNe~Ia, and all other sub-types, including the \uncIa, should not be considered.

In Fig.~\ref{fig:silvelmass} for those SNe Ia with \SiIIs\ velocity measurements around peak light, we do not see as many 91T-like SNe~Ia in the low-mass galaxies as in Fig.~\ref{fig:x1mass} for those with reliable light curves. A possible explanation for this could be that since the 91T-like events are brighter, they are possibly picked-up in earlier phases and the classifications were done before the maximum light phase range. 

\begin{figure}
\centering
\includegraphics[width=\columnwidth]{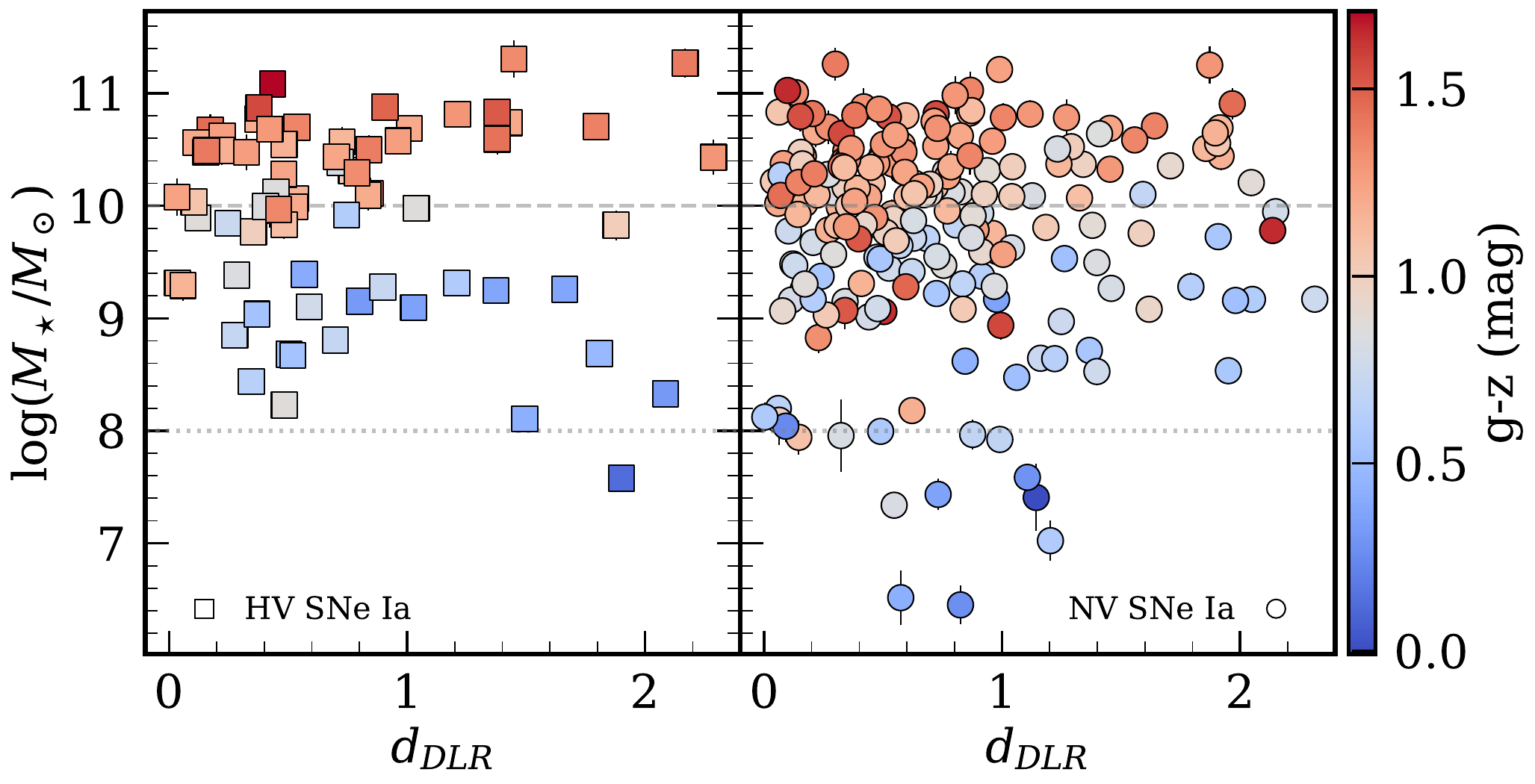}
\caption{The d$_{DLR}$ distribution of the `normal' SNe~Ia are presented as a function of stellar galaxy masses, color-mapped according to the global $g-z$ rest-frame colors. Circles and squares represent NV and HV SNe~Ia, respectively. Dashed horizontal line represent the separation of high mass and low mass galaxies at (log(\Mstar/\Msun) $=$ 10). There are 8 NV SNe~Ia with higher d$_{DLR}$ values that are excluded from the plot to provide a clearer comparison between the two subtypes within the lower d$_{DLR}$ ranges.} 
\label{fig:dDLRmassgzvel}
\end{figure}

\begin{figure*}
\minipage{0.5\textwidth}
  \includegraphics[width=\columnwidth]{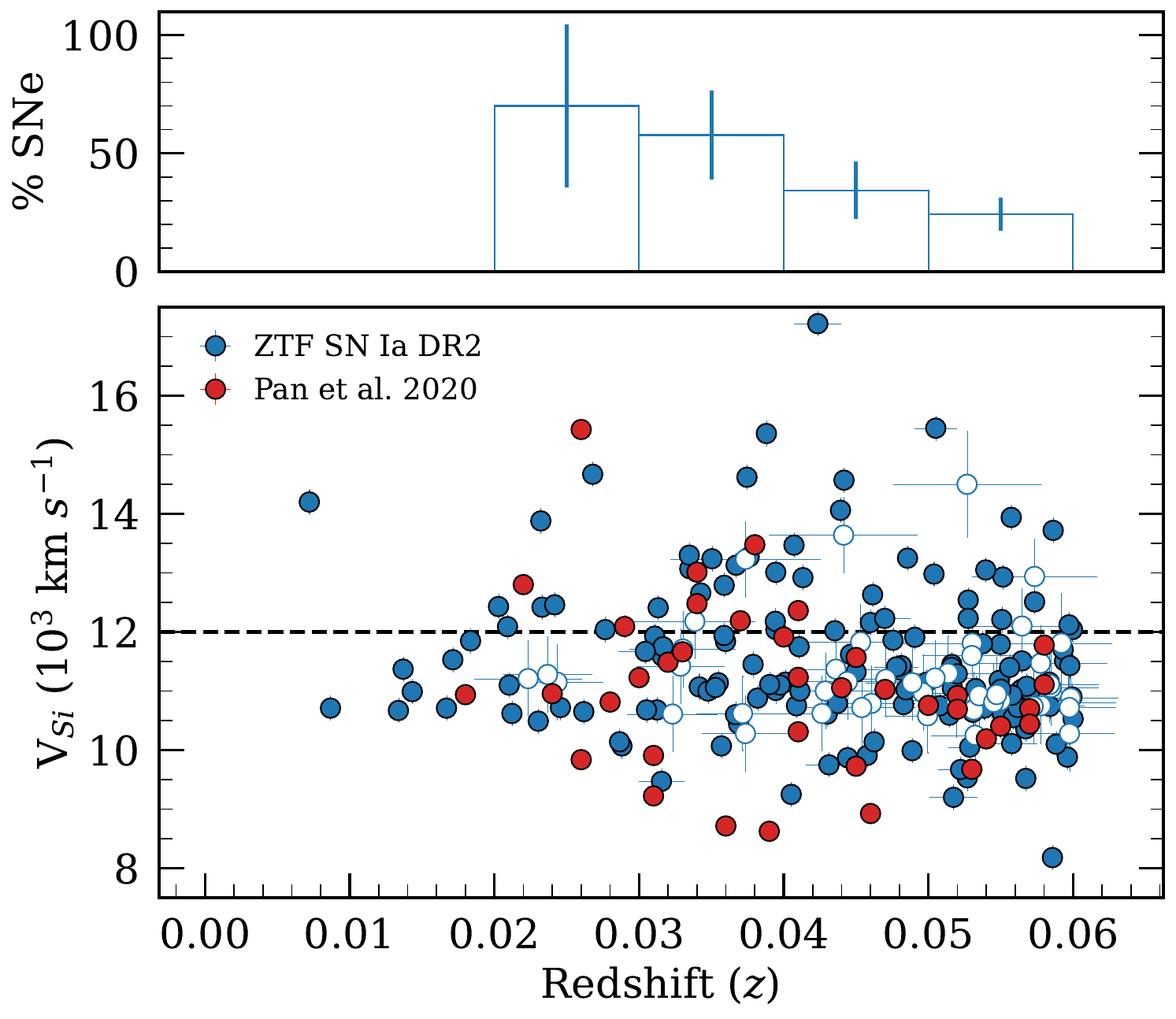}
\endminipage\hfill
\minipage{0.5\textwidth}
  \includegraphics[width=\columnwidth]{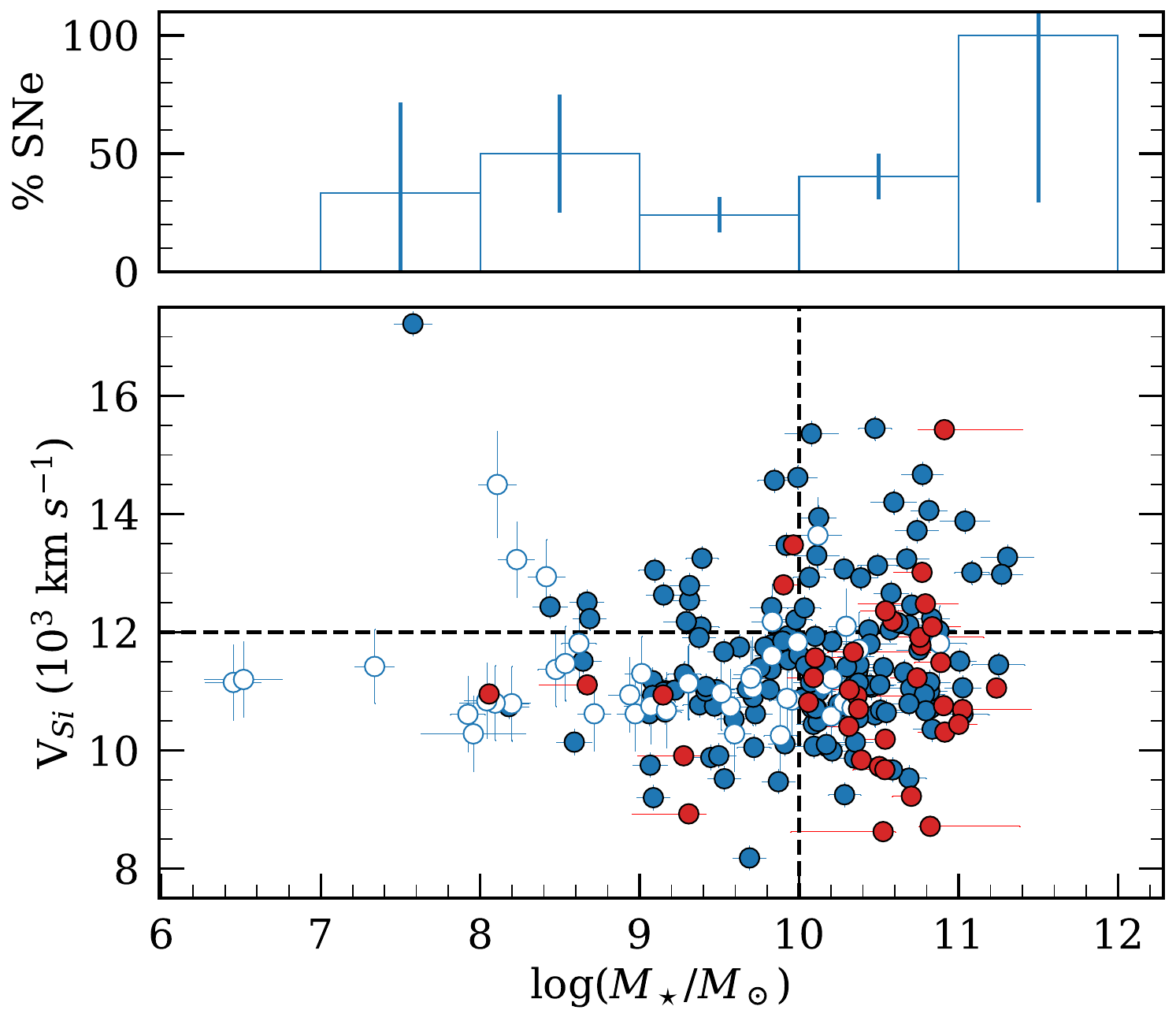}
\endminipage\hfill
\caption{\textbf{Left:} \SiIIs\ velocities plotted against the redshift. \textbf{Right:} \SiIIs\ velocities plotted against the stellar galaxy masses. In both plots, blue filled and unfilled circles represent the volume-limited ZTF DR2 SN Ia sample adjusted to match the selection criteria of P20 with known and template matched redshift sources within 3 days since peak brightness, respectively. Red circles represent the P20 sample with only PTF sources and matched to the redshift range of the volume-limited ZTF DR2 SN Ia sample, ($z$ $\leq$ 0.06). The horizontal black dashed line represents the value of \SiIIs\ velocity used to separate the sample into HV and NV SN Ia. The vertical dashed line represents the typical high-low host galaxy mass separation value. Histograms of the distributions are shown in the top panels for the corresponding plots underneath. In these plots, the unfilled bars represent the ratio of HV SNe~Ia to NV SNe Ia in the corresponding bins for the the volume-limited ZTF DR2 SN Ia sample.}
\label{fig:sil_comp}
\end{figure*}

The impact of host galaxy contamination on the measured \SiIIs\ velocities is not negligible \customcitecolor{red}{burgaz2024}, with decreases in the \SiIIs\ velocities of up to 500 \kms\ measured for SN Ia spectra when host galaxy contamination is increased. The d$_{DLR}$ (the offset of the SN location scaled by the galaxy size) can be used as a proxy for host galaxy contamination since more centrally located SNe Ia are more heavily impacted by the underlying galaxy light. In this work, we investigate the impact of host galaxy contamination through the d$_{DLR}$ measurements on the velocities of both the HV and NV SN~Ia samples. In Fig.~\ref{fig:dDLRmassgzvel}, we show the host galaxy stellar mass against the d$_{DLR}$ of each of the normal SNe Ia in our sample split into HV and NV events. The data points are color-coded based on the global rest-frame $g-z$ host colors (\md{Smith et al. in prep.}).

We do not identify any difference between the distribution of d$_{DLR}$ among the HV and NV sub-types (Fig.~\ref{fig:dDLRmassgzvel}). The ratio of HV to NV SNe~Ia in 0.5 $d_{DLR}$ bins remains the same up to $d_{DLR}$ $\sim$ 2, with a constant ratio of $\sim$ 30 per cent HV to NV events between $d_{DLR}$ of 0 and 2. A KS test for the d$_{DLR}$ distribution between the HV and NV SNe~Ia also yields a high p value of 0.85, meaning both classes will be affected similarly by any host galaxy contamination. Although there are no HV SNe~Ia with d$_{DLR}$ values exceeding 2.4, the sample includes 8 NV SNe~Ia with d$_{DLR}$ values above this threshold. It can also be seen that both HV SNe~Ia and NV SNe~Ia exhibit similar behavior in terms of rest-frame color distribution in both low and high mass host galaxies. There is only one HV SNe~Ia in the `low-mass' bin of our sample and it is with a higher value  of $d_{DLR} = 1.9$. The rest of the `low-mass' galaxies are with NV SNe~Ia at smaller $d_{DLR}$. However, within the NV SN~Ia sample, the `low-mass' and `intermediate-mass' samples have marginally higher mean $d_{DLR}$ of 0.89$\pm$0.18 and 0.94$\pm$0.09, respectively, than the mean $d_{DLR}$ of the `high-mass' counterpart (0.79$\pm$0.06). In summary, we interpret these results to mean that any difference in velocity between HV and NV subsamples is not driven by the impact of host galaxy contamination and this diversity is an intrinsic property of normal SNe Ia. 

\subsection{Comparison with other galaxy samples}
\label{sec:comppan}

The results of \cite{pan2014} and P20 suggest a trend where more massive galaxies host more HV SNe Ia compared to low-mass host galaxies. In this section, we compare the distribution of \SiIIs\ velocity versus stellar galaxy masses of our sample to those presented by P20, which used the galaxy stellar masses of \citet{pan2014}. Since our sample and theirs have different initial selection criteria, we first bring our sample down to a phase range of $-$3 d $\leq$ $t_0$ $\leq$ 3 d to match the phase range of P20, which drops the sample size to 186 SNe. 45 out 186 SNe in our sample have their redshifts estimated with \textsc{snid} template matching. In the P20 sample, only sources with spectroscopic redshifts were used. To investigate potential selection bias, we will conduct our analysis both by including and excluding the 45 template-matched sources. This will help determine if the inclusion of these sources affects our results. We disregard all non-PTF SNe~Ia to remove any selection bias from the sample of P20 and cut the SN Ia at higher redshift than the ZTF DR2 volume-limited redshift of $z$ $\leq$ 0.06, which drops the final sample of P20 to 37 SNe~Ia.

The resulting redshift distribution of our sample in terms of \SiIIs\ velocity is compared to that of P20 in the left hand panels of Fig.~\ref{fig:sil_comp}. The two distributions are found to be similar with a KS test between the two samples having a high p-value of 0.28 and 0.09 for the 45 template-matched sources excluded and included, respectively, suggesting in both cases the ZTF and P20 samples are likely to be coming from the same distribution. However, it can be seen that our sample has more HV SNe Ia at higher redshift than the P20 sample. We only have one HV SNe up to the redshift of 0.02 in our sample. Between the redshift of 0.02 to 0.03, the ratio of HV to NV SNe~Ia for the ZTF sample is $\sim$70 per cent. Similarly, between the redshift of 0.03 to 0.04, the ratio is $\sim$60. Above this, we start to have fewer HV SNe~Ia and the ratio changes to around $\sim$35 per cent for the redshift between 0.04 to 0.05, for the last redshift bin of 0.05 to 0.06, the ratio drops $\sim$24 per cent. The ratio indeed drops in the last bin, however it is still consistent with the previous bins within the uncertainties. A linear line fit to the data shows that the significance of the slope is inconsistent with zero slope at 2 $\sigma$ level. We note that this two-sigma significance is derived from the sample with $[-3:3]$ days phase range. The significance drops to 1.5 $\sigma$ for sample with $[-5:5]$ days phase range. Therefore, in this analysis we do not notice a significant change in the relative ratio of HV to NV SNe Ia, unlike P20. We note that our sample is limited to a redshift up to 0.06 and whether the current relative ratio of HV to NV SNe Ia holds at higher redshifts needs to be investigated.

In the right-hand panels of Fig.~\ref{fig:sil_comp}, the \SiIIs\ velocity against host stellar mass is shown for both samples. In the P20 sample (with only PTF SNe Ia and with $z \leq 0.06$), there are no HV SNe Ia below a host stellar mass of log(\Mstar/\Msun) of $\sim$9.9. In our sample, there are more HV SNe Ia in lower mass hosts compared to the P20 sample, with 28$\pm$7 per cent of SNe Ia in galaxies below 10$^{10}$ \Msun\ being HV compared to 40$\pm$9 per cent in galaxies above 10$^{10}$ \Msun. Our highest velocity SN Ia (SN 2018koy) is the event with a low stellar mass host. However, apart from this event, the SNe Ia with velocities above $\sim$14,000 \kms\ in our sample are mainly present in the higher mass hosts. 

\cite{Dettman2021} used data from the untargeted Foundation Supernova Survey \citep{foley2018} and divided their sample ($-$6 d $<$ $t_0$ $<$ 10 d) into four quadrants as having HV and NV split at $v_{\mathrm{Si}} < 11,800$ \kms\ in low and high mass galaxies split at log(\Mstar/\Msun) = 9.5 and found that the HV velocity SNe are found nearly exclusively in higher stellar mass (log(\Mstar/\Msun) > 9.5) hosts. However, they note that the significance of this result is dependent on where the split in host galaxy mass is made, if it is lowered to log(\Mstar/\Msun) = 9, then the result become insignificant. By following the same method we divide Fig.~\ref{fig:sil_comp} (right) into four quadrants. However, to evaluate the impact of various selection criteria on our results, we conducted our analysis considering each different criterion for splitting HV and NV velocities, masses, different phase ranges and the inclusion or exclusion of the template-matched sources, as outlined in Table \ref{tab:samp_sel_eff}. 

In our analysis with a HV and NV split at $v_{\mathrm{Si}} < 12,000$ \kms\ and low and high mass galaxies split at log(\Mstar/\Msun) = 10 with a phase range of $[-5,5]$ days with the template matched sources included, we have 109 NV and 40 HV SNe in high stellar mass and 110 NV SNe in low stellar mass. Hence, under the assumption of HV SNe Ia rate being same in low and high stellar masses, we predict 110(40/109) = 40$\pm$6 SNe in the low-mass HV quadrant, compared to the 34 SNe observed. Exclusion of template matched sources brings the predicted number to 23$\pm$5 SNe in the low-mass HV quadrant, compared to the 26 SNe observed. However, the sample with a smaller phase range, $[-3,3]$ days, gives 31$\pm$6 and 19$\pm$4 SNe predicted, compared to the 20 and 16 observed SNe for the template-matched sources included and excluded cases, respectively. 

Following the split criteria used in \cite{Dettman2021}, for the ZTF sample with $[-5,5]$ days phase range, we find that we have already more SNe observed than the predicted numbers in low-mass HV quadrant regardless of template-matched sources being included or excluded Table \ref{tab:samp_sel_eff}. The ZTF SN Ia sample with $[-3,3]$ days phase range also yields observed numbers that are already higher than or very close to the predicted values. In both split criteria cases, we note that higher phase ranges such as $[-5,5]$ days provides better results. Overall, we conclude that HV SNe Ia show similar rates of occurrence between high and low mass galaxies.

\begin{table}
\tiny
\centering
\caption{Sample selection effect on the number of HV SNe Ia in low-mass galaxies.}
\label{tab:samp_sel_eff}
\begin{tabular}{c c c c c }
\hline\\[-0.5em]

v$_{\text{\SiII}}$(6355)$^*$ & log(\Mstar/\Msun)$^\dagger$ & Phase & Redshift$^a$ & \# HV SN Ia$^b$ \\ 
 (x10$^{\text{3}} $ km s$^{-1}$) & & ($d$) & ($z$) & \\

\hline\\[-0.8em]
\hline\\[-0.5em]

12 & 10 & $[-5:5]$ & all & 34 (40$\pm$6) \\ 
12 & 10 & $[-5:5]$ & exc & 26 (23$\pm$5) \\ 
12 & 10 & $[-3:3]$ & all & 20 (31$\pm$6) \\ 
12 & 10 & $[-3:3]$ & exc & 16 (19$\pm$4) \\ 
11.8 & 9.5 & $[-5:5]$ & all & 29 (27$\pm$5) \\ 
11.8 & 9.5 & $[-5:5]$ & exc & 19 (15$\pm$4) \\ 
11.8 & 9.5 & $[-3:3]$ & all & 16 (18$\pm$4) \\ 
11.8 & 9.5 & $[-3:3]$ & exc & 12 (11$\pm$3) \\ 

\hline\\[-0.5em]

\end{tabular}

{\raggedright($^*$) The split value for the HV and NV SNe Ia. \par}
{\raggedright($^\dagger$) The split value for the low and high mass galaxies. \par}
{\raggedright($^a$) Sample with all redshift sources are shown as all and without those of \textsc{snid} template matched sources are shown as exc. \par}
{\raggedright($^b$) Number of HV SNe Ia in low-mass galaxies. The expected values, assuming that the rate of HV SNe Ia is the same irrespective of host galaxy stellar mass, are shown in parentheses, along with Poisson uncertainties. \par}

\end{table}

\cite{nugent2023} identified no trend between SN Ia \SiIIs\ velocity and host stellar mass but did suggest that they are more centrally concentrated. \cite{polin2019} suggested that HV SNe Ia may come from a population of sub-Chandrasekhar mass explosions.   P20 also measured gas-phase metallicities for a sub-sample of their events and speculated that increased metallicity may result in HV SNe Ia being formed. \cite{hantang2024} investigated the connection between velocity and age and found a possible correlation in the sense that HV SNe Ia may come from older environments but this result was low significance. Based on our results, we do not find any evidence for a difference in the stellar mass distribution of SNe Ia displaying HV \SiIIs\ features, compared to the NV sample. 

\begin{figure*}
\centering
\includegraphics[width=18cm]{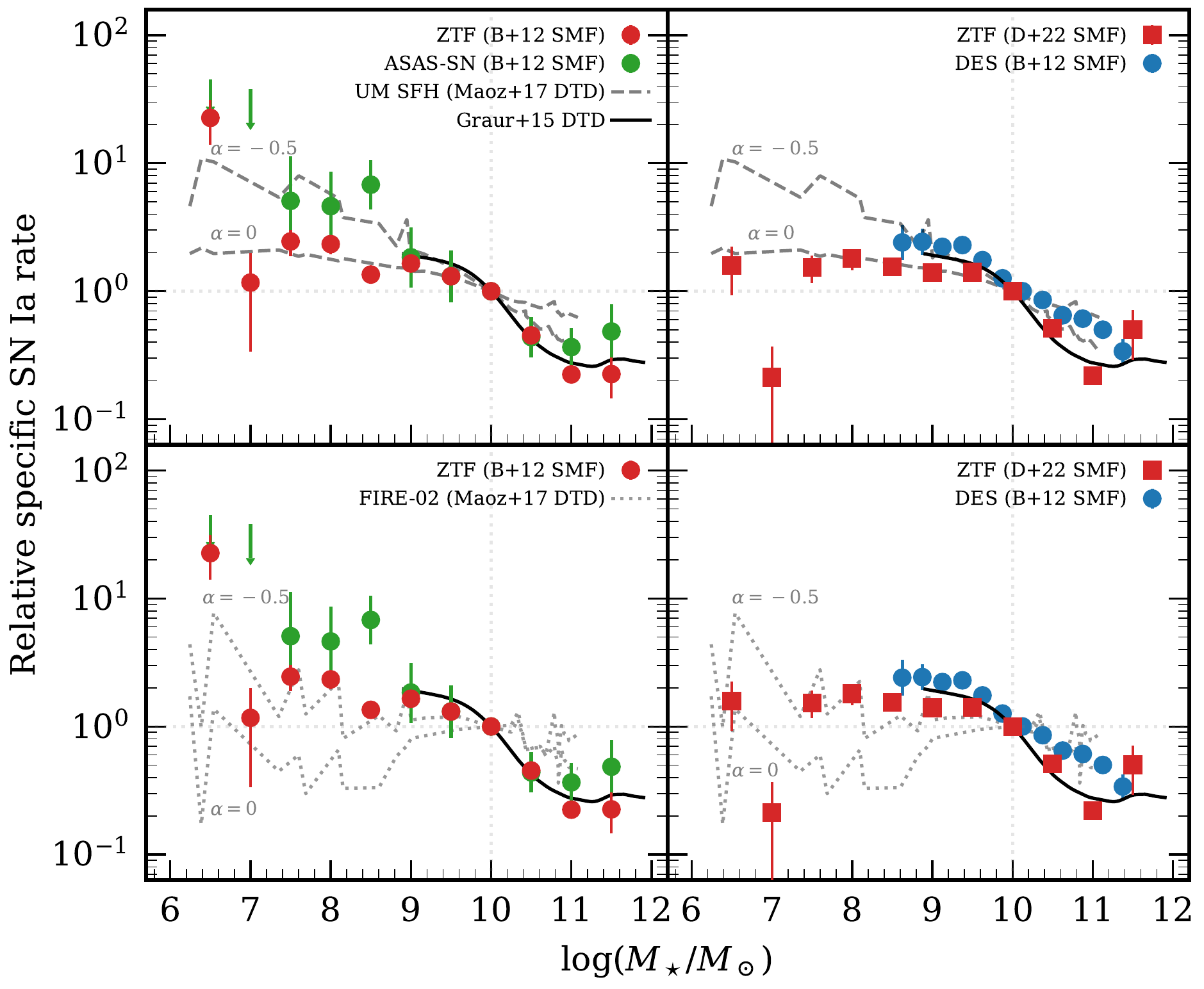}
\caption{SN Ia rate per unit stellar mass as function of stellar galaxy mass. As in \citet{brown2019} and \citet{gandhi2022}, all the rates are normalized to the rate at log(\Mstar/\Msun) $=$ 10. \textbf{Upper left:} The filled red dots represent the volume-limited ZTF SN Ia DR2 sample, using the SMF from \citet{baldry2012} as presented. Green dots represent the ASAS-SN volume-limited sample from the \citet{gandhi2022}, where they modified the SMF from \citet{Bell2003} used in \citet{brown2019} to that of \citet{baldry2012}. The rates obtained by applying the DTD from \citet{maoz2017} to semi-empirical galaxy SFHs from \citep[UniverseMachine, ][]{Behroozi2019}, as presented in \citet{gandhi2022}, are represented by the gray dashed lines (re-scaled to the ZTF data) where $\alpha$ = 0, indicating no metallicity dependence, and $\alpha$ = -0.5, a slight metallicity dependence as metallicity decreases, respectively.  \textbf{Upper right:} The filled squares represent the volume-limited ZTF SN Ia DR2 sample, using the SMF with the presented Schechter function from \citet{driver2022}. The DES sample from \citet{wiseman2021} is shown by blue points. The DTD from \citet{graur2015} is shown with the black line. \textbf{Lower left:} Same data as upper left panel, but with the rates from FIRE-2 simulations with the DTD applied from \citet{maoz2017} as used in \citet{gandhi2022} and presented with the gray dotted lines. \textbf{Lower right:} Same as upper right panel but with the FIRE-2 simulations.}
\label{fig:rate}
\end{figure*}

\subsection{Specific SN~Ia rate of the volume-limited ZTF DR2 SN Ia sample}
\label{sec:rates}

In this section, we calculate the SN~Ia rate per unit stellar mass for the volume-limited ZTF SN~Ia DR2 sample by combining our observed SN Ia rates with an estimate of the galaxy stellar mass function (SMF). Our sample covers the mass range from log(\Mstar/\Msun) of $\sim$6.5 to 11.5, with significantly more SNe Ia in the `low-mass' bins compared to previous studies, such as \cite{brown2019}. We note that the 6 hostless SNe identified in this study are not included in the rate analysis, since they all are calculated as upper limits and cannot be assigned to specific mass bins. 

In order to obtain the relative SN~Ia rate per unit stellar mass, the total number of SNe~Ia in each mass bin of width of log(\Mstar/\Msun) of  0.5 is divided by the integral of the stellar mass function over the width of that mass range. In this analysis as discussed below, we use the SMF from \cite{baldry2012} and the double-Schechter function presented in \cite{driver2022} to calculate the SMF. We do not attempt an absolute specific SN Ia rate but use the convention of \cite{brown2019} to show the relative rates normalized to the rate at log(\Mstar/\Msun) $=$ 10.

The ZTF SN Ia DR2 sample is defined as complete up to $z$ $\leq$ 0.06, but this is only valid for the bright 91T-like, bright transitional 99aa-like and normal SNe~Ia. \customcitecolor{red}{Dimitriadis2024} showed that fainter 91bg-like SNe~Ia are complete up to redshift $z$ $\leq$ 0.04. Even then, this is true only for the 91bg-like events down to an absolute magnitude of around $-$15 mag due the magnitude limitations of the ZTF survey. Hence, we are incomplete below this magnitude limit to 91bg-like events, although fainter 91bg-like events are not expected to be common \customcitepcolor{red}{Dimitriadis2024}. In this work, we introduce a correction to the volume-limited sample to account for the missing 91bg-like SNe Ia between redshifts of 0.04 and 0.06 based on the intrinsic rates calculated by \customcitecolor{red}{Dimitriadis2024}. We assume that that host galaxy stellar mass distribution of the missing events follow the same distribution as the observed sample of the missed events.

There are 75 91bg-like events out of 1523 SNe~Ia in the volume-limited ZTF DR2 sample, where 51 of the 91bg-like events are at $z$ $\leq$ 0.04. In order to find out how many 91bg-like events we miss, first we calculated the distribution of 91bg-like events in each mass bin, scaled to the total number of SNe~Ia. Then the same is done for the redshift range of 0.04 to 0.05 and 0.05 to 0.06 to find separate, more precise efficiencies as a function of mass \customcitepcolor{red}{Dimitriadis2024}. This correction introduces another 34 91bg-like events between 0.04 and 0.06 that were missed by ZTF due to its incompleteness for fainter 91bg-like SNe Ia. However, it is also important to point out that we might still be missing the faintest 91bg-like events, which could potentially increase the observed rate mainly for log(\Mstar/\Msun) $\geq$ 10, where they are most common.

\citet{gandhi2022} showed that assumed galaxy SMFs is the dominant sources of uncertainty in the specific SN~Ia rate and especially for the low mass galaxies, the SMF highly affects the SN Ia rate.  Previous studies have used a number of different estimates of the SMF. \cite{brown2019} used the SMF of \cite{Bell2003}, while \cite{gandhi2022} using the measured SMF from the low-redshift GAMA survey \citep{baldry2012}, which is steeper at lower mass. This SMF is considered complete down to a log(\Mstar/\Msun) of 8, with lower limits to log(\Mstar/\Msun) of 6.25. \citet{wiseman2021} for the higher redshift DES sample estimated their own SMF, which was in good agreement with the SMF of \cite{baldry2012}. More recently, the GAMA survey has released an updated sample, which extends their measured SMF values down to log(\Mstar/\Msun) of 6.75 with improved statistics \citep{driver2022}.  

We estimate the relative SN Ia rate per unit stellar mass for the ZTF data assuming two different methods of estimating the SMF. The first chosen method is similar to \cite{gandhi2022} using the slightly modified SMF of \cite{baldry2012} for the low-redshift GAMA survey and is also very similar to that used in DES \citep{wiseman2021}. The second method uses the double Schechter fit from the \cite{driver2022}, wich has measured SMF values down to log(\Mstar/\Msun) of 6.75. A steeper SMF at low stellar mass results in lower relative specific SN Ia rates at these masses. 

Fig.~\ref{fig:rate} shows the relative specific SN Ia rate for the volume-limited ZTF SN Ia DR2 sample (corrected for missed faint 91bg-like events) as a function of host galaxy stellar mass for the two methods of estimating the SMF, solid red points for the \cite{baldry2012} and red squares for the steeper double-Schechter fit of \cite{driver2022}. We observe a higher rate for the low mass galaxies with the SMF from \cite{baldry2012}, whereas the double-Schechter fit from \cite{driver2022} results a lower rate. Hence, the observed rates are highly dependent on the selected SMF. We notice a significant decrease in the specific SN Ia rate in the log(\Mstar/\Msun) = 7.0 bin - however, we believe this is due to the low number statistics where only two SNe are in the corresponding bin. For comparison, the log(\Mstar/\Msun) = 6.5 and 7.5 bins have 9 and 21 SNe, respectively.

The ASAS-SN volume limited sample of 113 SNe Ia at $z < 0.02$ \citep{brown2019} is also shown in Fig.~\ref{fig:rate}, assuming the SMF of \cite{baldry2012} used by \citet{gandhi2022}. We also show the DES absolute specific SN Ia rate of \citet{wiseman2021} for SNe Ia in the range of $0.2 < z < 0.6$, normalised to their rate at log(\Mstar/\Msun) $=$ 10 for comparison with our values. Looking first at the comparison of the rates in higher mass galaxies (log(\Mstar/\Msun) $\geq$10), we find the ZTF rates are slightly lower but consistent within the uncertainties to the ASAS-SN volume-limited sample \citep{brown2019,gandhi2022}. These are also consistent with the high mass (log(\Mstar/\Msun) > 10) rates from Lick Observatory SN Search as a function of stellar mass \citep{li2011}. We have made a correction to our volume-limited sample to account for missed 91bg-like SNe Ia in the range of $0.04 z < 0.06$. \cite{brown2019} noted that for SNe Ia fainter than $-$18 mag in the \textit{V} band they are incomplete. These fainter events have a strong preference to occur in higher mass hosts (see Fig.~\ref{fig:x1mass}) and so could result in a lower ASAS-SN rate in the highest mass host galaxies. However, we see the opposite trend with a slightly lower relative specific rate for the ZTF volume-limited sample in host galaxies with log(\Mstar/\Msun) $>$ 11. Again, we note that, if the ZTF rates are calculated with the double-Schechter fit, the relative specific rate for the log(\Mstar/\Msun) = 11 bin shows an increase, showing the effects coming from the choice of SMF.

Our rates are very similar to those of the ASAS-SN sample from log(\Mstar/\Msun) of 9 to 10. However, for the lower masses the ASAS-SN volume-limited sample show generally higher rates. Below log(\Mstar/\Msun) of 8, we have used two estimates of the SMF. The ZTF data shown with solid red points use \cite{baldry2012} SMF, while the the ASAS-SN data \citep{gandhi2022} uses their own conversion from \cite{Bell2003} to \cite{baldry2012} SMF. However, the modified SMF is still very close to the \cite{baldry2012} values. Hence, the differences observed between the two samples are highly likely to be driven by the measured rates of the SNe Ia in those mass bins, expect for the significant drop observed at the log(\Mstar/\Msun) = 7 bin for the ZTF data, where only two SNe exist. The lowest ASAS-SN sample bins are upper limits since in their sample they only had one SNe with log(\Mstar/\Msun) $<$ 7.25.

In each mass bin where the DES data exists, we see that both the ZTF and ASAS-SN samples show lower rates in general than those of the DES sample. Since a more strict $x_1$ cut is applied in \cite{wiseman2021}, there is about 10 per cent of their sample removed, due to the cut from the $x_1$ being below $-$3, which is usually populated by 91bg-like events.

\citet{gandhi2022} investigated how the specific SN Ia rate depends on the metallicity by testing different parameters for the metallicity in the SN Ia DTD from \cite{maoz2017} combined with the star formation histories (SFH) of the galaxy simulations, FIRE-2 \citep{hopkins2018} and UniverseMachine \citep[UM;][]{Behroozi2019}. They found that a single power-law Z$^\alpha$ metallicity dependence to the SN Ia DTD, with $\alpha$ = $-$0.5 to $-$1, is the closest to their observed rates.  In the top panels of Fig.~\ref{fig:rate}, the ZTF data in the low-mass galaxies closely follow the UM models combined with a DTD with no metallicity dependence ($\alpha = 0$). However, an increased metallicity dependence is needed to follow the observed trend in the high mass galaxies. The ASAS-SN date with its higher SN Iarates, on average, in the lower mass bins is more consistent with a metallicity dependence but is based on a smaller sample with a number of upper limits.  The choice of SMF also contributes to the observed differences between the model and the data. On average, the \cite{driver2022} SMF matches the models better than that of \cite{baldry2012}. 

In the bottom panels of Fig.~\ref{fig:rate}, we compare our ZTF data using the two different SMF to the FIRE-2 simulations combined with the metallcity dependent DTD. Using the FIRE-2 SFH, compared to the UM SFH, results in the ZTF data better matching the models with $\alpha = -0.5$, suggesting a metallicity dependence in the lower mass host galaxies. In summary, the lower and generally flatter specific rates of SNe Ia at lower stellar mass seen for the ZTF sample compared to ASAS-SN sample, combined with the most up-to-date \cite{driver2022} SMF (right-hand panels of Fig.~\ref{fig:rate}), suggest if that there is a metallicity dependence to the observed relative specific SN Ia rate is small and perhaps zero. 

We also included the rate simulations from \citet{graur2015} in our comparisons, which extend down to log(\Mstar/\Msun) $\simeq$ 9. We observe that our rates are generally consistent with these simulations, although the simulated rates tend to predict slightly higher values than those observed in the actual ZTF data for galaxies with log(\Mstar/\Msun) $\leq$ 10.

\section{Conclusions}
\label{sec:conclusions}

In this work, we used sub-samples of 1523 SNe~Ia from the volume-limited ZTF DR2 SN Ia sample to explore the relation between the velocity of the \SiIIs\ feature, SALT2 light-curve stretch $x_1$ and stellar galaxy properties of SNe~Ia. 
\begin{enumerate}
      \item We visually inspected six hostless SNe~Ia in our sample to identify possible host galaxies nearby. Upon investigation 4 out of 6 hostless SN Ia were found to be near the Galactic plane (within 15 degrees) and experiencing higher extinction. One hostless SN Ia, while not being within 15 degrees from the Galactic plane, was still close enough for extinction to be the reason for a lack of detected host. The remaining one hostless SN Ia is further away from the Galactic plane with low extinction and we found an extremely faint, diffuse source with no detection in the PS1 images. 
      \item We investigated the distribution of SALT2 stretch $x_1$ and stellar galaxy masses for different subtypes in our sample. A general trend of brighter SNe~Ia being in the lower-mass galaxies is observed, agreeing with previous studies. In the low and intermediate mass  galaxies, the distribution of both normal SN Ia and Ia-unclear subtype remains similar for each mass bin. Bright 91T-like SNe~Ia and 99aa-like transitional SNe~Ia have a higher distribution in the lower mass regions. In the `low-mass' (log(\Mstar/\Msun) $\leq$ 8) galaxies, around 80 per cent of the SNe~Ia are normal SNe~Ia, around 10 per cent are Ia-unclear and remaining 10 per cent are 91T-like and 99aa-like SNe~Ia. 
      \item Using a volume-limited SN Ia sample, we found that HV SNe~Ia do not necessarily favor high-mass galaxies but are also found in low-mass and intermediate mass galaxies, even without including the redshift sources from \textsc{snid} matches. We conclude that the missing HV SNe~Ia in previous samples mainly relates to the sample size and/or a selection bias towards observing higher mass galaxies.
      \item No correlations have been found between the rest-frame $g-z$ host galaxy color, as well as the SN to host galaxy distance scaled with the size of the corresponding galaxies between HV and NV SNe Ia, indicating both sub-types should be experiencing similar host galaxy contamination.
      \item Following the work of \citet{gandhi2022}, we calculated the specific SN Ia rate for the volume-limited ZTF SN Ia DR2 sample and compared with other surveys, as well as a DTD model. Overall, we find our rates to be closely in agreement with the ASAS-SN volume-limited sample within the uncertainties. In the low mass galaxies, our data differs from the ASAS-SN data and show a more constant rate. By comparison with literature SFH and a metallicity-dependent DTD, we find that our sample is most consistent with little to no metallicity dependence of the specific SN Ia rate.  
   \end{enumerate} 

In this work, we highlight the significance of low-mass galaxies and the value of large, well-characterized samples of such galaxies, highlighting the need for even larger samples in future studies. A forthcoming paper will explore the relationship between metallicity and low-mass host galaxies, specifically focusing on systems with stellar masses log(\Mstar/\Msun) $\leq$ 8. 

\begin{acknowledgements}
We thank Or Graur for providing simulation data for SNe Ia rates. UB, KM, GD, MD and JHT are supported by the H2020 European Research Council grant no. 758638. LH is funded by the Irish Research Council under grant number GOIPG/2020/1387. T.E.M.B. acknowledges financial support from the Spanish Ministerio de Ciencia e Innovaci\'on (MCIN), the Agencia Estatal de Investigaci\'on (AEI) 10.13039/501100011033, and the European Union Next Generation EU/PRTR funds under the 2021 Juan de la Cierva program FJC2021-047124-I and the PID2020-115253GA-I00 HOSTFLOWS project, from Centro Superior de Investigaciones Cient\'ificas (CSIC) under the PIE project 20215AT016, and the program Unidad de Excelencia Mar\'ia de Maeztu CEX2020-001058-M.
This project has received funding from the European Research Council (ERC) under the European Union's Horizon 2020 research and innovation program (grant agreement n 759194 - USNAC).
L.G. acknowledges financial support from AGAUR, CSIC, MCIN and AEI 10.13039/501100011033 under projects PID2020-115253GA-I00, PIE 20215AT016, CEX2020-001058-M, and 2021-SGR-01270.
Y.-L.K. has received funding from the Science and Technology Facilities Council [grant number ST/V000713/1].
This work has been supported by the research project grant “Understanding the Dynamic Universe” funded by the Knut and Alice Wallenberg Foundation under Dnr KAW 2018.0067 and the  {\em Vetenskapsr\aa det}, the Swedish Research Council, project 2020-03444.

Based on observations obtained with the Samuel Oschin Telescope 48-inch and the 60-inch Telescope at the Palomar Observatory as part of the Zwicky Transient Facility project. ZTF is supported by the National Science Foundation under Grants No. AST-1440341 and AST-2034437 and a collaboration including current partners Caltech, IPAC, the Weizmann Institute of Science, the Oskar Klein Center at Stockholm University, the University of Maryland, Deutsches Elektronen-Synchrotron and Humboldt University, the TANGO Consortium of Taiwan, the University of Wisconsin at Milwaukee, Trinity College Dublin, Lawrence Livermore National Laboratories, IN2P3, University of Warwick, Ruhr University Bochum, Northwestern University and former partners the University of Washington, Los Alamos National Laboratories, and Lawrence Berkeley National Laboratories. Operations are conducted by COO, IPAC, and UW. SED Machine is based upon work supported by the National Science Foundation under Grant No. 1106171. The ZTF forced-photometry service was funded under the Heising-Simons Foundation grant 12540303 (PI: Graham). SED Machine is based upon work supported by the National Science Foundation under Grant No. 1106171. The ZTF forced-photometry service was funded under the Heising-Simons Foundation grant 12540303 (PI: Graham). This work was supported by the GROWTH project funded by the National Science Foundation under Grant No 1545949 \citep{kasliwal2019}. Fritz \citep{vander2019, coughlin2020} is used in this work. The Gordon and Betty Moore Foundation, through both the Data-Driven Investigator Program and a dedicated grant, provided critical funding for SkyPortal. 

\end{acknowledgements}
\bibliographystyle{aa}
\bibliography{aanda} 

\end{document}